\documentstyle[aps, epsf, psfig]{revtex}
\begin{document}
\draft
\title{Covariant amplitudes for mesons}
\author{Jie-Jie Zhu\footnote{Electronic address: jjzhu@ustc.edu.cn}, 
Mu-Lin Yan }
\address{
CCAST(World Lab), P. O. Box 8730,
Beijing, 100080, P. R. China\\
and Center for Fundamental Physics, USTC,
Hefei, Anhui, 230026, P. R. China\footnote{Permanent address.}
}
\date{\today}
\maketitle
\begin{abstract}
We show how to construct covariant amplitudes for processes involving
higher spins in this paper.
First we give the explicit expressions of 
Rarita-Schwinger wave functions and propagators for bosons with spins, 
then kinematic singularity free 3-leg effective vertexes are derived 
and given in a list. Equivalence relations are worked out to get these
independent vertexes.
Constraints of space reflection symmetry and boson symmetry are considered 
and shown in a explicit way. Some helicity amplitudes for two-body decays in 
center of frame are calculated.
Finally the covariant helicity amplitudes for the process 
$a_1\rightarrow \pi^+\pi^+\pi^-$ are constructed to illustrate
how to include background (1PI) amplitudes. Both background amplitudes and
resonance amplitudes are needed to give reliable descriptions to high
energy reactions.
\end{abstract}

\section{Introduction}
Model independent amplitudes are needed to analyze high energy experimental 
data. Such amplitudes are usually written in terms of helicity formalism
first proposed by Jacob and Wick~\cite{JacobH59,ChouH58,BermanH65,WickH62},
or tensor formalism by Zemach~\cite{ZemachTA64,ZemachTA65}, in a 
non-covariant form. Recently Chung~\cite{ChungTA93} and Filippini 
{\it et al.}~\cite{FilippiniTA95} emphasized on the importance of covariance 
to get reliable results. Chung gave some examples of two-body decays with 
their amplitudes calculated in center of mass frame in  his 
works~\cite{ChungTA93,ChungSF71} on covariant tensor formalism. 
The case of spin-$J$ $\rightarrow$ spin-$j$ + spin-$0$ with 
$J, j \leq 2$ has been discussed in detail in Ref.~\cite{FilippiniTA95}.

We construct covariant amplitudes in a view of S-matrix approach. We call a 
S-matrix (maybe off shell) after stripping off external lines 
(wave functions) effective vertexes. Effective vertexes are related to 
Green functions by LSZ reduction formulations~\cite{LSZ55,BogoliubovAFT}, 
which can be divide into one-particle irreducible (1PI) parts and one-particle 
reducible parts. The former ones are often called backgrounds and the later 
resonances. One should repeat the process to divide those sub-vertexes 
connected by (full) propagators in resonance parts, until arrives at 3-leg 
vertexes that can not be separated. 

These effective vertexes (with three or more legs) should be constructed from
the four-momenta of outer legs and isotropic tensors of Lorentz group. We use
wave functions satisfying Rarita-Schwinger conditions~\cite{RaritaWF41}
in this paper. This specific choice of wave functions will not introduce any 
model dependence, since we can change our results into any other 
representations via basis transformations. Information on the structure of 
particles is contained in effective vertexes. The general form of an effective
vertex for bosons are tensors to be constructed from 
$p_i^\mu$ ($i=1,2,\cdots.)$ and 
$g^{\mu\nu}, \varepsilon^{\alpha\beta\gamma\delta}$. Following the assumption 
of maximal analyticity~\cite{StappAF62,HaraKS64,StappKS67}, such vertex 
functions are free of kinematic singularities 
(K. S.)~\cite{HaraKS64,WangKS66,TannoudjiKS68,BardeenKS68,BrownKS71}. 
However, one should be careful when writing down independent effective 
vertexes. Since there are redundant components to be removed by 
Rarita-Schwinger conditions, some seemingly independent terms are related 
by equivalence relations~\cite{ScadronVF68}. We work out equivalence 
relations and give a general list of 3-leg effective vertexes. Some special
cases of them have already been given by Scadron~\cite{ScadronVF68}.

Additional symmetries give constraints on the form factors in effective 
vertexes. Space reflection symmetry demands effective vertexes being tensors
or pseudo-tensors, depending on spin-parities of their outer legs. The ratio
of form factors in tensor parts and those in pseudo-tensor parts can be taken
as a parameter of parity violation. For 3-leg effective vertexes involving
two spin-$0$ particles, some processes are forbidden.

If an effective vertex is connected to two identical bosons, it must satisfy
boson symmetry. There are kinematic zeros in form factors that should be 
factored out. When both of the two bosons are on shell, the ``anti-symmetric''
parts of the 3-leg effective vertex vanish. For the case of one particle is on
shell while another is off shell, 
the contribution from ``anti-symmetric'' parts is not independent 
from background terms, since the form factors in these parts
contribute a factor which eliminates the denominator of the propagator for
the off shell particle. We give a list of 3-leg effective vertexes, with
all these cases considered.

Covariant amplitudes in different reference frame are related by Wigner 
rotations~\cite{WignerLG39}. We calculate the amplitudes for some two-body 
decays in center of mass (CM) frame. The relation between amplitudes in
laboratory frame and those in CM frame is derived.

Background amplitudes should not be neglected in order to give a full 
description to a reaction, and to get reliable information from 
data~\cite{Zhua23pi}. 1PI amplitudes will not give flat distributions 
if there are particles with non-zero spins. 
The covariant helicity amplitudes for $a_1\rightarrow \pi^+\pi^+\pi^-$ 
are constructed as a demonstration.

Our present work is partly based on Ref.\cite{ZhuPHD}. In Sec.~\ref{sec:wf}
we give a brief introduction to wave functions and propagators. 3-leg effective
vertexes are listed in Sec.~\ref{sec:3v}. Sec.~\ref{sec:sr} and 
Sec.~\ref{sec:bs} are devoted to the constraint of space reflection symmetry 
and boson symmetry. Two-body decays are considered in Sec.~\ref{sec:2bd}. 
In Sec.~\ref{sec:bk} we discuss the process $a_1\rightarrow \pi^+\pi^+\pi^-$.

\section{Wave functions}\label{sec:wf}
Let $L(p)$ be a Lorentz transformation,
    \begin{equation}\label{eq:lrtz}
    p^\mu={L^\mu}_\nu(p)k^\nu.
    \end{equation}
The standard momentum for a mass-$W$ particles is $(k^\mu)=(W;\ \roarrow{0})$. 
The space-time metric we use is
$(g^{\mu\nu})={\rm diag}\{1,-1,-1,-1\}.$
Now define one-particle states as~\cite{WeinbergQFT1}
    \begin{equation}\label{eq:qsdef}
    |p,\sigma\rangle =U(L(p))|k,\sigma\rangle
    \end{equation}
with $U(L(p))$ the unitary representation of $L(p)$ in Hilbert
space. It satisfies
    \begin{equation}
    \hat{p}^\mu|p,\sigma\rangle=p^\mu|p,\sigma\rangle .
    \end{equation}

We can choose the orthonormal condition to be
    \begin{equation}
    \langle p^{'}, \sigma^{'} | p, \sigma \rangle =
    (2 \pi)^3(2 p^0) \delta(\roarrow{p}^{'} - \roarrow{p})
    \delta_{\sigma^{'}\sigma}.
    \end{equation}
Under a Lorentz transformation $\Lambda$,
    \begin{equation}\label{eq:qswr}
    U(\Lambda)| p,\sigma \rangle = \sum\limits_{\sigma^{'}}
      D_{\sigma^{'}\sigma}(W(\Lambda,p))|
       \Lambda p, \sigma^{'} \rangle,
    \end{equation}
where
    \begin{equation}
    W(\Lambda,p)
        \equiv L^{-1}(\Lambda p)\Lambda L(p)
    \end{equation}
is the Wigner rotation~\cite{WignerLG39} and
$\{D_{\sigma^{'}\sigma}\}$ furnishes a representation of $SO(3).$

If we define the Lorentz transformation in Eq.~(\ref{eq:qsdef})
to be a pure Lorentz boost
    \begin{equation}\label{eq:lc}
    \begin{array}{rcl}
    L(p)&=&L(\roarrow{p})\\
        &\equiv&R(\varphi,\theta,0)L_{z}
            (|\roarrow{p}|)R^{-1}(\varphi,\theta,0),
    \end{array}
    \end{equation}
we obtain canonical states. Here $R(\varphi,\theta,0)$ is the
rotation that takes $z$-axis to the direction of $\roarrow{p}$, and the boost
$L_{z}(|\roarrow{p}|)$ takes the four-momentum
$(k^\mu)=(W;\ \roarrow{0})$ to
$\left( \sqrt{W^2+\roarrow{p}^2};\ 0,\ 0,\ |\roarrow{p}|\right)$.
For a particle of spin-$j$, $\sigma\sim(j,m)$.
For canonical states,
Eq.~(\ref{eq:qswr}) becomes
    \begin{equation}
    U(\Lambda)|\roarrow{p},j,m\rangle=\sum\limits_{m^{'}}
      D_{m^{'} m}^j(L^{-1}(\roarrow{\Lambda}p)\Lambda L(\roarrow{p}))
      |\roarrow{\Lambda}p,j,m^{'}\rangle.
    \end{equation}
$ D_{m^{'} m}^j $ is the ordinary $D$-function. Especially, under
a rotation $R$,
    \begin{equation}\label{eq:rot1}
    U(R)|\roarrow{p},j,m\rangle=\sum\limits_{m^{'}}
      D_{m^{'} m}^j(R)
      |\roarrow{R}p,j,m^{'}\rangle.
    \end{equation}

Defining the Lorentz transformation in another way leads to
helicity states~\cite{JacobH59}:
    \begin{equation}\label{eq:lh}
    \begin{array}{rcl}
    L(p)&=&L(\roarrow{p})R^{-1}(\varphi,\theta,0)\\
        &\equiv&R(\varphi,\theta,0)L_{z}(|\roarrow{p}|).
    \end{array}
    \end{equation}
We have
    \begin{equation}\label{eq:lrt}
    U(\Lambda)|\roarrow{p},j,\lambda\rangle=\sum\limits_{\lambda^{'}}
        D_{\lambda^{'}\lambda}^j
            (L^{-1}(\roarrow{\Lambda}p)\Lambda L(\roarrow{p}))
    |\roarrow{\Lambda}p,j,\lambda^{'}\rangle
    \end{equation}
and
    \begin{equation}\label{eq:rot2}
    U(R)|\roarrow{p},j,\lambda\rangle=
      |\roarrow{R}p,j,\lambda\rangle.
    \end{equation}

The two kinds of definitions are related to each other,
    \begin{equation}
    |\roarrow{p},j,\lambda\rangle_{helicity}
    =\sum\limits_m D^j_{m\lambda}(\varphi,\theta,0)
        |\roarrow{p},j,m\rangle_{canonical}.
    \end{equation}

Quantum states in terms of  creation and annihilation operators read:
    \begin{equation}
    |\roarrow{p}, j, \sigma\rangle=
      \sqrt{(2 \pi)^3 2 p^0}a^\dagger(\roarrow{p}, j, \sigma)|0\rangle,
    \end{equation}
where $|0\rangle$ is the vacuum state. We use ``$\sigma$'' to mean that the
relation holds for both helicity states and canonical states.
Quantum fields for spin-$j$ bosons are constructed as~\cite{WeinbergQFT1}
\begin{equation}
\phi_{\mu_1\mu_2...\mu_j}=\\
\int\frac{d^3 p}{\sqrt{\left(2\pi\right)^3 2 p^0}}
\{
\sum\limits_\sigma e_{\mu_1\mu_2...\mu_j}(\roarrow{p}, j, \sigma)
	a(\roarrow{p}, j, \sigma) e^{-ip\cdot x} 
+ \sum\limits_\sigma e^{\ast}_{\mu_1\mu_2...\mu_j}(\roarrow{p}, j, \sigma)
	{a^c}^\dagger (\roarrow{p}, j, \sigma)e^{ip\cdot x}
\},
\end{equation}
with ${a^c}^\dagger$ the annihilation operator of the antiparticle, and 
\begin{equation}
U(\Lambda,a)\phi_{\mu_1\mu_2...\mu_j}U^{-1}(\Lambda,a)=
  {\Lambda_{\mu_1}}^{\nu_1} {\Lambda_{\mu_2}}^{\nu_2}... 
   {\Lambda_{\mu_j}}^{\nu_j} \phi_{\nu_1\nu_2...\nu_j}(\Lambda x +a).
\end{equation}
$e_{\mu_1\mu_2...\mu_j}$ is the wave function in
momentum space satisfying~\cite{WeinbergQFT1}
\begin{equation}\label{eq:wltr}
 {\Lambda_{\mu_1}}^{\nu_1} {\Lambda_{\mu_2}}^{\nu_2}... 
 {\Lambda_{\mu_j}}^{\nu_j}
 e_{\nu_1\nu_2...\nu_j}(\roarrow{p}, j, \sigma)
=\sum\limits_{\sigma^{'}}D_{\sigma^{'} \sigma}(W(\Lambda,p))
  e_{\mu_1\mu_2...\mu_j}(\roarrow{p}, j, \sigma^{'}),
\end{equation}
so its definition is
\begin{equation}
 e_{\mu_1\mu_2...\mu_j}(\roarrow{p}, j, \sigma) = 
	{\Lambda_{\mu_1}}^{\nu_1} {\Lambda_{\mu_2}}^{\nu_2} ... 
	{\Lambda_{\mu_j}}^{\nu_j}
	 e_{\mu_1\mu_2...\mu_j}(\roarrow{0}, j, \sigma).
\end{equation}

From the following infinitesimal generators of the Lorentz group
    \begin{equation}
    \begin{array}{ccc}
    \left({{J_1}^\mu}_\nu\right)= 
      \left(
      \begin{array}{cccc}
      0 & 0 & 0 & 0\\
      0 & 0 & 0 & 0\\
      0 & 0 & 0 & -i\\
      0 & 0 & i & 0
      \end{array}
      \right), &
    \left({{J_2}^\mu}_\nu\right)= 
      \left(
      \begin{array}{cccc}
      0 & 0 & 0 & 0\\
      0 & 0 & 0 & i\\
      0 & 0 & 0 & 0\\
      0 & -i & 0 & 0
      \end{array}
      \right), &
    \left({{J_3}^\mu}_\nu\right)= 
      \left(
      \begin{array}{cccc}
      0 & 0 & 0 & 0\\
      0 & 0 & -i & 0\\
      0 & i & 0 & 0\\
      0 & 0 & 0 & 0
      \end{array}
      \right), \\
    \left({{K_1}^\mu}_\nu\right)= 
      \left(
      \begin{array}{cccc}
      0 & i & 0 & 0\\
      i & 0 & 0 & 0\\
      0 & 0 & 0 & 0\\
      0 & 0 & 0 & 0
      \end{array}
      \right), &
    \left({{K_2}^\mu}_\nu\right)= 
      \left(
      \begin{array}{cccc}
      0 & 0 & i & 0\\
      0 & 0 & 0 & 0\\
      i & 0 & 0 & 0\\
      0 & 0 & 0 & 0
      \end{array}
      \right), &
    \left({{K_3}^\mu}_\nu\right)= 
      \left(
      \begin{array}{cccc}
      0 & 0 & 0 & i\\
      0 & 0 & 0 & 0\\
      0 & 0 & 0 & 0\\
      i & 0 & 0 & 0
      \end{array}
      \right),
    \end{array}
    \end{equation}
one can obtain 
\begin{equation}
\left({L_z\left(|\roarrow{p}|\right)^\mu}_\nu\right) 
\equiv  e^{-i \alpha K_3}  
 =  \left(
\begin{array}{cccc}
\cosh \alpha & 0 & 0 & \sinh \alpha \\
0 	     & 1 & 0 & 0 \\
0 	     & 0 & 1 & 0 \\
\sinh \alpha & 0 & 0 & \cosh \alpha
\end{array}
\right) 
= \left(
\begin{array}{cccc}
E/W 		& 0 & 0 & |\roarrow{p}|/W \\
0 		& 1 & 0 & 0 \\
0 		& 0 & 1 & 0 \\
|\roarrow{p}|/W & 0 & 0 & E/W
\end{array}
\right),
\end{equation}
\begin{equation}
\left(R\left(\varphi, \theta, 0\right){^\mu}_\nu \right) \equiv 
\left({
\left(
e^{-i \varphi J_3}e^{-i \theta J_2}
\right)^\mu}_\nu
\right) =
\left(
\begin{array}{cccc}
1 & 0 			     & 0 	     & 0 \\
0 & \cos \theta \cos \varphi & -\sin \varphi & \sin \theta \cos \varphi \\
0 & \cos \theta \sin \varphi & \cos \varphi  & \sin \theta \sin \varphi \\
0 & - \sin \theta 	     & 0	     & \cos \theta 
\end{array}
\right).
\end{equation}
Choose wave functions at rest frame to be 
    \begin{eqnarray}
	\left(e^\mu(\roarrow{0},0)\right) & = &
		\left(
		\begin{array}{c}
		0\\0\\0\\1
		\end{array}
		\right), \\
	\left(e^\mu(\roarrow{0},\pm 1)\right) & = &
		\mp \frac{1}{\sqrt{2}}
		\left(
		\begin{array}{c}
		0\\1\\ \pm i \\ 0
		\end{array}
		\right),
    \end{eqnarray}
we get the familiar explicit expressions of spin-$1$ canonical wave 
functions($E$ is the energy of the particle)
    \begin{equation}\label{eq:wfc}
    \begin{array}{rcl}
    \left(e_c^\mu(\roarrow{p},0)\right) & = &
          \left(
          \begin{array}{c}
            \frac{|\roarrow{p}|}{W}\cos\theta\\
            \frac{1}{2}\left(\frac{E}{W}-1\right)
                \sin 2\theta\cos\varphi\\
            \frac{1}{2}\left(\frac{E}{W}-1\right)
                \sin 2\theta\sin\varphi\\
            \frac{1}{2}\left(\frac{E}{W}-1\right)(1+\cos 2\theta)+1
          \end{array}  
          \right),\\
    \left(e_c^\mu(\roarrow{p},\pm 1)\right) & = &
          \mp\frac{1}{\sqrt{2}}\left(
          \begin{array}{c}
            \frac{|\roarrow{p}|}{W}\sin\theta e^{\pm i\varphi}\\
            \left(\frac{E}{W}-1\right)\sin^2\theta\cos\varphi
                    e^{\pm i\varphi}+1\\
            \left(\frac{E}{W}-1\right)\sin^2\theta\sin\varphi
                    e^{\pm i\varphi}\pm 1\\
            \left(\frac{E}{W}-1\right)\cos\theta\sin\theta
                    e^{\pm i\varphi}
          \end{array}  
          \right)
    \end{array}
    \end{equation}
and spin-$1$ helicity wave functions
\begin{equation}\label{eq:wfh}
    \begin{array}{rcl}
    \left(e_h^\mu(\roarrow{p},0)\right) & = &
          \left(
          \begin{array}{c}
            \frac{|\roarrow{p}|}{W}\\
            \frac{E}{W}\sin \theta\cos\varphi\\
            \frac{E}{W}\sin \theta\sin\varphi\\
            \frac{E}{W}\cos \theta
          \end{array}  
          \right),\\
    \left(e_h^\mu(\roarrow{p},\pm 1)\right) & = &
          \frac{1}{\sqrt{2}}\left(
          \begin{array}{c}
            0\\
            \mp\cos\theta\cos\varphi +i\sin\varphi\\
            \mp\cos\theta\sin\varphi -i\cos\varphi\\
            \pm\sin\theta
          \end{array}  
          \right).
    \end{array}
\end{equation}

Wave functions for higher integral spin particles can be defined recurrently by
\begin{equation}\label{eq:wfj}
      e_{\mu_1\mu_2\cdots\mu_j}(\roarrow{p},j,\sigma)
    =\sum\limits_{\sigma^{'}_{j-1},\sigma_j}
      (j-1,\sigma^{'}_{j-1};1,\sigma_j|j,\sigma)
      e_{\mu_1\mu_2\cdots\mu_{j-1}}(\roarrow{p},j-1,\sigma_{j-1}^{'})
      e_{\mu_j}(\roarrow{p},\sigma_j).
\end{equation}
Using the following C-G coefficient relation
\begin{equation}
\begin{array}{rl}
 & \sum\limits_{\sigma^{'}_3, \sigma^{'}_4, \cdots, \sigma^{'}_n}
  (j_1,\sigma_1;j_2,\sigma_2|j_1+j_2,\sigma^{'}_3)
  (j_1+j_2,\sigma^{'}_3;k_3,\sigma_3|j_1+j_2+j_3,\sigma^{'}_4)\cdots\\
 & \ \ \ \times
  (j_1+j_2+\cdots+j_{n-1},\sigma^{'}_n;
        j_n,\sigma_n|j_1+j_2+\cdots+j_n,
        \sigma^{'}_n+\sigma_n)\\
=&\left[\prod\limits_{i=1}^{n}
    \frac{(2j_i)!}{(j_i+\sigma_i)! (j_i-\sigma_i)!}\right]^\frac{1}{2}
   \left\{\frac{\left[\sum\limits^n_{i=1}(j_i+\sigma_i)\right]!
            \left[\sum\limits^n_{i=1}(j_i-\sigma_i)\right]!}
           {\left(2\sum\limits_{i=1}^n j_i\right)!}
   \right\}^\frac{1}{2},
\end{array}
\end{equation}
we find
\begin{equation}
\begin{array}{rl}
 &  e_{\mu_1\mu_2\cdots\mu_j}(\roarrow{p},j,\sigma) \\
=& \sum\limits_{\sigma_1, \sigma_2,\cdots, \sigma_j}
   \left\{ \frac{2^j(j+\sigma)!(j-\sigma)!}
      {(2j)!\prod\limits^j_{i=1}[(1+\sigma_i)!(1-\sigma_i)!]}
   \right\}^{\frac{1}{2}}
    \delta_{\sigma_1+\sigma_2+\cdots+\sigma_j,\sigma}
  e_{\mu_1}(\roarrow{p},\sigma_1)
                e_{\mu_2}(\roarrow{p},\sigma_2)\cdots
     e_{\mu_j}(\roarrow{p},\sigma_j).
\end{array}
\end{equation}
The above expression is equivalent to that given by Scadron~\cite{ScadronVF68}
and Chung~\cite{ChungHA95}, since they come from the same definition 
of Eq.~(\ref{eq:wfj}). 

$e_{\mu_1\mu_2\cdots\mu_j}(\roarrow{p},j,\sigma)$ satisfies
Rarita-Schwinger conditions: space-like, symmetric and traceless:
\begin{eqnarray}					\label{eq:rs1}
p^{\mu_1}  e_{\mu_1\mu_2...\mu_j}(\roarrow{p}, j, \sigma) & = & 0,\\ 
							\label{eq:rs2}
 e_{\mu_1...\mu_k...\mu_l...\mu_j}(\roarrow{p}, j, \sigma) & = &
   e_{\mu_1...\mu_l...\mu_k...\mu_j}(\roarrow{p}, j, \sigma),\\ 
							\label{eq:rs3}
 g^{\mu_1\mu_2} e_{\mu_1\mu_2...\mu_j}(\roarrow{p}, j, \sigma) & = & 0.
\end{eqnarray}

Spin projection operator is~\cite{FronsdalSP58} 
\begin{equation}\label{eq:prj}
\begin{array}{rl}
& {\cal P}^{(j)}_{\mu_1 \mu_2 ... \mu_j ; \nu_1 \nu_2 ... \nu_j} \\
 \equiv &  
	\sum\limits_{\sigma} e_{\mu_1\mu_2...\mu_j}(\roarrow{p},j,\sigma)
   e^\ast_{\nu_1\nu_2...\nu_j}(\roarrow{p},j,\sigma) \\
 = & \frac{1}{(j!)^2}
	\sum\limits_{\begin{array}{l}
			P\left\{ \mu_1 \mu_2 ... \mu_j \right\} \\
		     	P\left\{ \nu_1 \nu_2 ... \nu_j \right\}  
		     \end{array}
		    }
\left(
 \prod\limits^{j}_{i=1}\tilde{g}_{\mu_i \nu_i}+
	a_{1}^{(j)} \tilde{g}_{\mu_1 \mu_2} \tilde{g}_{\nu_1 \nu_2} 
		\prod\limits^{j}_{i=3}\tilde{g}_{\mu_i \nu_i}
 + ...
\right . \\ &
 + \left\{
	\begin{array}{ll}
	 \left .
	a_{j/2}^{(j)}\prod\limits_{i=1}^{j/2}\left(
	   \tilde{g}_{\mu_{2i-1}\mu_{2i}}\tilde{g}_{\nu_{2i-1}\nu_{2i}}
					     \right)
	\right), & {\rm for\; even}\; j, \\
	\left. 
	a_{(j-1)/2}^{(j)} \tilde{g}_{\mu_j\nu_j}
	 \prod\limits_{i=1}^{(j-1)/2}\left(
	   \tilde{g}_{\mu_{2i-1}\mu_{2i}}\tilde{g}_{\nu_{2i-1}\nu_{2i}}
				     \right)
	\right), & {\rm for\; odd}\; j;
	\end{array}
    \right . 
\end{array}
\end{equation}
where
\begin{equation}
a_K^{(j)} = \frac{(-1)^K j! } {2^K K! (j-2K)!}
	\frac{1} {(2j-1)(2j-3)...(2j-2K+1)},
\end{equation}
and
\begin{equation}
\tilde{g}_{\mu\nu} = -g_{\mu \nu} + \frac{{p_\mu} {p_\nu}} {W^2}
\end{equation}
is the spin-$1$ projection operator.
Spin projection operators are useful when we sum over final 
(or initial) particles' spins. 
They also serve as numerators of propagators~\cite{WeinbergQFT1}.
From Eq.~{(\ref{eq:prj})}, the first five projection operators read
\begin{eqnarray}
{\cal P}^{(1)}_{\mu \nu} &  = & -g_{\mu \nu} + \frac{{p_\mu} {p_\nu}} {W^2} ,\\
{\cal P}^{(2)}_{\mu_1 \mu_2 ;\nu_1 \nu_2} &  = &
	\frac{1}{2}(\tilde{g}_{\mu_1 \nu_1} \tilde{g}_{\mu_2 \nu_2} +
		\tilde{g}_{\mu_2 \nu_1} \tilde{g}_{\mu_1 \nu_2}) -
	\frac{1}{3}\tilde{g}_{\mu_1 \mu_2}\tilde{g}_{\nu_1 \nu_2}, 
\end{eqnarray}
\begin{equation}
\begin{array}{rl}
{\cal P}^{(3)}_{\mu_1 \mu_2 \mu_3 ;\nu_1 \nu_2 \nu_3} & = \\ 
    +\frac{1}{6}\sum\limits_{P\{\nu_1,\nu_2,\nu_3\}} &
      \tilde{g}_{\mu_1 \nu_1} \tilde{g}_{\mu_2 \nu_2} \tilde{g}_{\mu_3 \nu_3}\\
    -\frac{1}{30}\sum\limits_{P\{\nu_1,\nu_2,\nu_3\}} &
	(\tilde{g}_{\mu_1 \mu_2}\tilde{g}_{\nu_1 \nu_2}\tilde{g}_{\mu_3\nu_3}
	+\tilde{g}_{\mu_1 \nu_1}\tilde{g}_{\mu_2 \mu_3}\tilde{g}_{\nu_2\nu_3}
	+\tilde{g}_{\mu_1 \mu_3}\tilde{g}_{\nu_1 \nu_3}\tilde{g}_{\mu_2\nu_2}),
\end{array}
\end{equation}
\begin{equation}
\begin{array}{rl}
{\cal P}^{(4)}_{\mu_1 \mu_2 \mu_3 \mu_4; \nu_1 \nu_2 \nu_3 \nu_4} & = \\
    +\frac{1}{24}\sum\limits_{P\{\nu_1,\nu_2,\nu_3,\nu_4\}} & 
	\tilde{g}_{\mu_1\nu_1} \tilde{g}_{\mu_2\nu_2} \tilde{g}_{\mu_3\nu_3}
		\tilde{g}_{\mu_4\nu_4} \\
    -\frac{1}{168}\sum\limits_{P\{\nu_1,\nu_2,\nu_3,\nu_4\}} &
	(\tilde{g}_{\mu_1\mu_2}\tilde{g}_{\mu_3\nu_1}\tilde{g}_{\mu_4\nu_2}+
 	 \tilde{g}_{\mu_2\mu_3}\tilde{g}_{\mu_1\nu_1}\tilde{g}_{\mu_4\nu_2}+
	 \tilde{g}_{\mu_1\mu_3}\tilde{g}_{\mu_2\nu_1}\tilde{g}_{\mu_4\nu_2}+\\
&  	 \tilde{g}_{\mu_1\mu_4}\tilde{g}_{\mu_2\nu_1}\tilde{g}_{\mu_3\nu_2}+
	 \tilde{g}_{\mu_2\mu_4}\tilde{g}_{\mu_1\nu_1}\tilde{g}_{\mu_3\nu_2}+
	 \tilde{g}_{\mu_3\mu_4}\tilde{g}_{\mu_1\nu_1}\tilde{g}_{\mu_2\nu_2}
	)
	\tilde{g}_{\nu_3\nu_4} \\
    +\frac{1}{840}\sum\limits_{P\{\nu_1,\nu_2,\nu_3,\nu_4\}} &
	( \tilde{g}_{\mu_1\mu_2}\tilde{g}_{\mu_3\mu_4}+
	 \tilde{g}_{\mu_1\mu_3}\tilde{g}_{\mu_2\mu_4}+
	 \tilde{g}_{\mu_1\mu_4}\tilde{g}_{\mu_2\mu_3}
	)
	\tilde{g}_{\nu_1\nu_2}\tilde{g}_{\nu_3\nu_4},
\end{array}
\end{equation}
\begin{equation}
\begin{array}{rl}
{\cal P}^{(5)}_{\mu_1\mu_2\mu_3\mu_4\mu_5; \nu_1\nu_2\nu_3\nu_4\nu_5} & = \\
    +\frac{1}{120}\sum\limits_{P\{\nu_1,\nu_2,\nu_3,\nu_4,\nu_5\}} & 
	\tilde{g}_{\mu_1\nu_1}\tilde{g}_{\mu_2\nu_2}\tilde{g}_{\mu_3\nu_3}
		\tilde{g}_{\mu_4\nu_4}\tilde{g}_{\mu_5\nu_5} \\
    -\frac{1}{1080}\sum\limits_{P\{\nu_1,\nu_2,\nu_3,\nu_4,\nu_5\}} &
	(\tilde{g}_{\mu_1\mu_2}\tilde{g}_{\mu_3\nu_1}\tilde{g}_{\mu_4\nu_2}
		\tilde{g}_{\mu_5\nu_3}
	+\tilde{g}_{\mu_1\mu_3}\tilde{g}_{\mu_2\nu_1}\tilde{g}_{\mu_4\nu_2}
		\tilde{g}_{\mu_5\nu_3}\\
&       +\tilde{g}_{\mu_1\mu_4}\tilde{g}_{\mu_2\nu_1}\tilde{g}_{\mu_3\nu_2}
		\tilde{g}_{\mu_5\nu_3}
	+\tilde{g}_{\mu_1\mu_5}\tilde{g}_{\mu_2\nu_1}\tilde{g}_{\mu_3\nu_2}
		\tilde{g}_{\mu_4\nu_3}\\
& 	+\tilde{g}_{\mu_2\mu_3}\tilde{g}_{\mu_1\nu_1}\tilde{g}_{\mu_4\nu_2}
		\tilde{g}_{\mu_5\nu_3}
	+\tilde{g}_{\mu_2\mu_4}\tilde{g}_{\mu_1\nu_1}\tilde{g}_{\mu_3\nu_2}
		\tilde{g}_{\mu_5\nu_3}\\
& 	+\tilde{g}_{\mu_2\mu_5}\tilde{g}_{\mu_1\nu_1}\tilde{g}_{\mu_3\nu_2}
		\tilde{g}_{\mu_4\nu_3}
	+\tilde{g}_{\mu_3\mu_4}\tilde{g}_{\mu_1\nu_1}\tilde{g}_{\mu_2\nu_2}
		\tilde{g}_{\mu_5\nu_3}\\
& 	+\tilde{g}_{\mu_3\mu_5}\tilde{g}_{\mu_1\nu_1}\tilde{g}_{\mu_2\nu_2}
		\tilde{g}_{\mu_4\nu_3}
	+\tilde{g}_{\mu_4\mu_5}\tilde{g}_{\mu_1\nu_1}\tilde{g}_{\mu_2\nu_2}
		\tilde{g}_{\mu_3\nu_3}
	)\tilde{g}_{\nu_4\nu_5}\\
    +\frac{1}{7560}\sum\limits_{P\{\nu_1,\nu_2,\nu_3,\nu_4,\nu_5\}} &
	(\tilde{g}_{\mu_1\mu_2}\tilde{g}_{\mu_3\mu_4}\tilde{g}_{\mu_5\nu_1}
	+\tilde{g}_{\mu_1\mu_3}\tilde{g}_{\mu_2\mu_4}\tilde{g}_{\mu_5\nu_1}
	+\tilde{g}_{\mu_1\mu_4}\tilde{g}_{\mu_2\mu_3}\tilde{g}_{\mu_5\nu_1}\\
&	+\tilde{g}_{\mu_1\mu_2}\tilde{g}_{\mu_3\mu_5}\tilde{g}_{\mu_4\nu_1}
	+\tilde{g}_{\mu_1\mu_3}\tilde{g}_{\mu_2\mu_5}\tilde{g}_{\mu_4\nu_1}
	+\tilde{g}_{\mu_1\mu_5}\tilde{g}_{\mu_2\mu_3}\tilde{g}_{\mu_4\nu_1}\\
&	+\tilde{g}_{\mu_1\mu_2}\tilde{g}_{\mu_4\mu_5}\tilde{g}_{\mu_3\nu_1}
	+\tilde{g}_{\mu_1\mu_4}\tilde{g}_{\mu_2\mu_5}\tilde{g}_{\mu_3\nu_1}
	+\tilde{g}_{\mu_1\mu_5}\tilde{g}_{\mu_2\mu_4}\tilde{g}_{\mu_3\nu_1}\\
&	+\tilde{g}_{\mu_1\mu_3}\tilde{g}_{\mu_4\mu_5}\tilde{g}_{\mu_2\nu_1}
	+\tilde{g}_{\mu_1\mu_4}\tilde{g}_{\mu_3\mu_5}\tilde{g}_{\mu_2\nu_1}
	+\tilde{g}_{\mu_1\mu_5}\tilde{g}_{\mu_3\mu_4}\tilde{g}_{\mu_2\nu_1}\\
&	+\tilde{g}_{\mu_2\mu_3}\tilde{g}_{\mu_4\mu_5}\tilde{g}_{\mu_1\nu_1}
	+\tilde{g}_{\mu_2\mu_4}\tilde{g}_{\mu_3\mu_5}\tilde{g}_{\mu_1\nu_1}
	+\tilde{g}_{\mu_2\mu_5}\tilde{g}_{\mu_3\mu_4}\tilde{g}_{\mu_1\nu_1}
	)\tilde{g}_{\nu_2\nu_3}\tilde{g}_{\nu_4\nu_5}.
\end{array}
\end{equation}

Now we can state Feynman rules for bosons as:
(1) Every incoming particle or incoming antiparticle contributes a factor 
of $e_{\mu_1 \mu_2 ... \mu_j}(\roarrow{p}, j , \sigma)$. 
(2) Every outgoing particle or outgoing antiparticle contributes a factor 
of $e^{\ast}_{\mu_1 \mu_2 ... \mu_j}(\roarrow{p}, j , \sigma)$. 
(3) For each spin-$j$ internal line, include a factor~\cite{WeinbergQFT1}
\begin{equation}\label{eq:prop}
i \frac{ {\cal P}^{(j)}_{\mu_1 \mu_2 ... \mu_j ; \nu_1 \nu_2 ... \nu_j} } 
	{p^2 -W^2 + i \epsilon}.
\end{equation}
Note that we have dropped constants unnecessary for amplitude analysis.
The denominator of Eq.~{(\ref{eq:prop})} is often changed to 
Breit-Wigner form as an approximation to the full propagator:
\begin{equation}\label{eq:bw}
i \frac{ {\cal P}^{(j)}_{\mu_1 \mu_2 ... \mu_j ; \nu_1 \nu_2 ... \nu_j} } 
	{p^2 -W^2 + i \Gamma W},
\end{equation}
where $\Gamma$ is the width of the particle. The width is either 
determined from experiments or from the chain approximation of 
theoretical models.

The problem left now is how to write down effective vertexes. 
We will give a list of 3-leg effective vertexes in next section. 

\section{Three-leg effective vertexes}\label{sec:3v}
The effective vertex $\Gamma$ should be constructed from momentums and 
isotropic tensors of the Lorentz group. Free indexes left after contraction
of $p_i^\mu , g^{\mu \nu} , \varepsilon^{\alpha \beta \gamma \delta}$
are the Lorentz indexes of $\Gamma$. 
Here $p_i^\mu$ ($i = 1, 2, ... , n$) is the four-momenta for
particle $i$ and $\varepsilon^{\alpha \beta \gamma \delta}$ 
the antisymmetric tensor. 
Not all possible constructions are independent since we have 
conservation of energy and momentum
\begin{equation}
p_1^\mu + p_2^\mu + ... + p_n^\mu = 0
\end{equation}
and Rarita-Schwinger conditions. 

The antisymmetric tensor has the property
\begin{equation}\label{eq:2eps}
\varepsilon_{\mu \nu \alpha \beta} 
	\varepsilon_{\mu^{'} \nu{'} \alpha^{'} \beta^{'}}
= -\det	\left\{
\begin{array}{cccc}
g_{\mu \mu^{'}} & g_{\nu \mu^{'}} & 
	g_{\alpha \mu^{'}} & g_{\beta \mu^{'}} \\
g_{\mu \nu^{'}} & g_{\nu \nu^{'}} & 
	g_{\alpha \nu^{'}} & g_{\beta \nu^{'}} \\
g_{\mu \alpha^{'}} & g_{\nu \alpha^{'}} &
	 g_{\alpha \alpha^{'}} & g_{\beta \alpha^{'}} \\
g_{\mu \beta^{'}} & g_{\nu \beta^{'}} & 
	g_{\alpha \beta^{'}} & g_{\beta \beta^{'}}
\end{array}
	\right\},
\end{equation}
so products of $\varepsilon_{\mu \nu \alpha \beta}$ can be 
absorbed into other terms K. S. freely.

Let's move on to the case of effective vertexes with only three 
legs.

\vspace{2cm}
\begin{figure}[htb]
\centerline{\psfig{figure=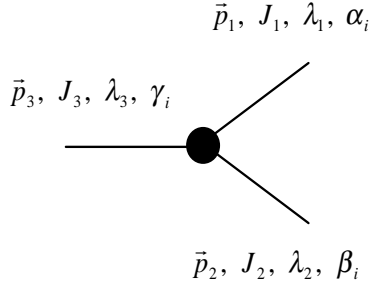,height=1.6in,width=2in}}
\caption{A three-leg effective vertex}\label{fig:3leg}
\end{figure}

As shown in Fig.~\ref{fig:3leg}, the four-momenta of the three
particles are $p_1$, $p_2$ and $p_3$. The spins are $J_1$,
$J_2$ and $J_3$, and helicities being $\lambda_1$, $\lambda_2$ 
and $\lambda_3$. The Lorentz indexes for these
particles are $(\alpha_1, \alpha_2, ..., \alpha_{J_1})$,
$(\beta_1, \beta_2, ..., \beta_{J_2})$ and 
$(\gamma_1, \gamma_2, ..., \gamma_{J_3})$, so the effective
vertex $\Gamma$ has $J_1 + J_2 + J_3$ indexes.
 
The antisymmetric tensor $\varepsilon^{\alpha \beta \gamma \delta}$ 
in a three-leg vertex should contract with at least one four-momenta 
since wave functions are symmetric.

We define
\begin{eqnarray}
A_1^{\alpha \beta \gamma} & \equiv & p_{1 \mu} 
	\varepsilon^{\mu \alpha \beta \gamma}, \\
A_2^{\alpha \beta \gamma} & \equiv & p_{2 \mu} 
	\varepsilon^{\mu \alpha \beta \gamma}, \\
Q^{\mu \nu} & \equiv & p_{1 \alpha} p_{2 \beta}
	\varepsilon^{\alpha \beta \mu \nu}.
\end{eqnarray}

In effective vertexes involving higher spin particles, 
some seemingly independent terms we write down are in fact not
independent. There are equivalence relations among them. These equivalence
relations come from Eq.~{(\ref{eq:2eps})} and the following
identity~\cite{ScadronVF68}
\begin{eqnarray}
\varepsilon^{\alpha\beta\gamma\delta}g^{\mu\nu}
& = & \varepsilon^{\mu\beta\gamma\delta}g^{\alpha\nu}
+\varepsilon^{\alpha\mu\gamma\delta}g^{\beta\nu}
+\varepsilon^{\alpha\beta\mu\delta}g^{\gamma\nu}
+\varepsilon^{\alpha\beta\gamma\mu}g^{\delta\nu}.
\end{eqnarray}
We can find\footnote{
Three of these equivalence relations, Eq.~{(\ref{eq:scadron1})},
Eq.~{(\ref{eq:scadron2})} and Eq.~{(\ref{eq:scadron3})}, 
have been listed in Ref.~\cite{ScadronVF68}.}
\begin{eqnarray}\label{eq:scadron1}
Q^{\alpha_1 \gamma_1} p_1^{\beta_1} & 
\simeq & p_1^2 A_2^{\alpha_1 \beta_1 \gamma_1} 
	-(p_1 \cdot p_2)A_1^{\alpha_1 \beta_1 \gamma_1}
	+Q^{\alpha_1 \beta_1} p_1^{\gamma_1} , \\ \label{eq:scadron2}
Q^{\beta_1 \gamma_1} p_2^{\alpha_1} & 
\simeq & - (p_1 \cdot p_2)A_2^{\alpha_1 \beta_1 \gamma_1} 
	+p_2^2 A_1^{\alpha_1 \beta_1 \gamma_1}
	+Q^{\alpha_1 \beta_1} p_1^{\gamma_1},
\end{eqnarray}
\begin{eqnarray}
Q^{\alpha_1 \beta_1} g^{\alpha_2 \gamma_1} & 
\simeq & A_1^{\alpha_1 \beta_1 \gamma_1} p_2^{\alpha_2}
	+Q^{\alpha_1 \gamma_1} g^{\alpha_2 \beta_1},\\
Q^{\alpha_1 \beta_1} g^{\beta_2 \gamma_1} & 
\simeq &
-A_2^{\alpha_1 \beta_1 \gamma_1} p_1^{\beta_2}
	-Q^{\beta_1 \gamma_1} g^{\alpha_1 \beta_2} ,
\end{eqnarray}
\begin{equation}\label{eq:scadron3}
Q^{\alpha_1 \gamma_1} g^{\beta_1 \gamma_2}   
-Q^{\beta_1 \gamma_1} g^{\alpha_1 \gamma_2} 
\simeq
	A_1^{\alpha_1 \beta_1 \gamma_1} p_1^{\gamma_2} +
	A_2^{\alpha_1 \beta_1 \gamma_1} p_1^{\gamma_2},
\end{equation}
\begin{equation}
\begin{array}{rl}
& g^{\alpha_1 \gamma_1}p_1^{\beta_1}p_1^{\beta_2}p_1^{\gamma_2}p_2^{\alpha_2}
- g^{\beta_1 \gamma_1}p_1^{\beta_2}p_1^{\gamma_2}p_2^{\alpha_1}p_2^{\alpha_2}\\
\simeq &
 g^{\alpha_1\beta_1}p_1^{\beta_2}p_1^{\gamma_1}p_1^{\gamma_2}p_2^{\alpha_2}
+[(p_1\cdot p_2)^2 - p_1^2 p_2^2]
	g^{\alpha_1\beta_1}g^{\alpha_2\gamma_1}g^{\beta_2\gamma_2}\\ &
-\frac{1}{2}p_2^2 g^{\alpha_1\gamma_1}g^{\alpha_2\gamma_2}
	p_1^{\beta_1}p_1^{\beta_2}
-\frac{1}{2}p_1^2 g^{\beta_1\gamma_1}g^{\beta_2\gamma_2}
	p_2^{\alpha_1}p_2^{\alpha_2}\\ &
-\frac{1}{2}(p_1+p_2)^2 g^{\alpha_1\beta_1}g^{\alpha_2\beta_2}
	p_1^{\gamma_1}p_1^{\gamma_2}
-(p_1\cdot p_2)g^{\alpha_1\gamma_1}g^{\beta_1\gamma_2}
	p_1^{\beta_2}p_2^{\alpha_2}\\ &
+[(p_1\cdot p_2)+p_2^2]g^{\alpha_1\beta_1}g^{\alpha_2\gamma_1}
	p_1^{\beta_2}p_1^{\gamma_2}
-[(p_1\cdot p_2)+p_1^2]g^{\alpha_1\beta_1}g^{\beta_2\gamma_1}
	p_1^{\gamma_2}p_2^{\alpha_2},
\end{array}
\end{equation}
\begin{equation}
A_1^{\alpha_1\beta_1\gamma_1}p_2^{\alpha_2}g^{\beta_2\gamma_2}+
A_2^{\alpha_1\beta_1\gamma_1}g^{\alpha_2\gamma_2}p_1^{\beta_2}
\simeq 
- g^{\alpha_1\beta_1}Q^{\alpha_2\gamma_1}g^{\beta_2\gamma_2}
- g^{\alpha_1\beta_1}Q^{\beta_2\gamma_1}g^{\alpha_2\gamma_2},
\end{equation}
\begin{equation}
\begin{array}{rl}
& A_1^{\alpha_1\beta_1\gamma_1}p_2^{\alpha_2}p_1^{\beta_2}p_1^{\gamma_2}
+ A_2^{\alpha_1\beta_1\gamma_1}p_2^{\alpha_2}p_1^{\beta_2}p_1^{\gamma_2}\\
\simeq &
\frac{1}{2}p_1^2 A_2^{\alpha_1\beta_1\gamma_1}
	p_2^{\alpha_2}g^{\beta_2\gamma_2} -
\frac{1}{2}(p_1\cdot p_2)A_1^{\alpha_1\beta_1\gamma_1}
	p_2^{\alpha_2}g^{\beta_2\gamma_2} \\
& +
\frac{1}{2}(p_1\cdot p_2)A_2^{\alpha_1\beta_1\gamma_1}
	g^{\alpha_2\gamma_2}p_1^{\beta_2} -
\frac{1}{2}p_2^2 A_1^{\alpha_1\beta_1\gamma_1}
	g^{\alpha_2\gamma_2}p_1^{\beta_2} \\
& +
\frac{1}{2}[(p_1\cdot p_2)-p_1^2]g^{\alpha_1\beta_1}
	A_2^{\alpha_2\beta_2\gamma_1}p_1^{\gamma_2} \\
& +
\frac{1}{2}[(p_1\cdot p_2)-p_2^2]g^{\alpha_1\beta_1}
	A_1^{\alpha_2\beta_2\gamma_1}p_1^{\gamma_2} \\
& -
g^{\alpha_1\beta_1}Q^{\alpha_2\beta_2}
	p_1^{\gamma_1}p_1^{\gamma_2}.
\end{array}
\end{equation}

The symbol ``$\simeq$''  means that the left hand side and the right hand side
of the equation are equal only when they are contracted with wave functions 
satisfying Eqs.~{(\ref{eq:rs1}-\ref{eq:rs3})}.
The left hand side of these equations can be absorbed into those terms on the
right hand side without introducing kinematic singularities.

A K. S. free effective vertex is written as
\begin{equation}
\Gamma = \Gamma_+ + \Gamma_-,
\end{equation}
where $\Gamma_+$($\Gamma_-$) is the tensor(pseudo-tensor) part of the
vertex. We sort the three particles in ascending order of their spins,
i.e., $J_1 \leq J_2 \leq J_3$.
After considering these equivalence relations, one finds:

\begin{itemize}

\item $(J_1, J_2, J_3) = (0, 0, j)$, with $j \leq 0$

\begin{equation}
\Gamma_+ = c_1 \left(
p_1^{\gamma_1}p_1^{\gamma_2}\cdots p_1^{\gamma_j}
\right),
\end{equation}

\begin{equation}\label{eq:00j-}
\Gamma_- = 0;
\end{equation}

\item $(J_1, J_2, J_3) = (0, j, j^{'})$, with $1 \leq j \leq j^{'}$

\begin{equation}
\begin{array}{rcl}
\Gamma_+ & = &
\; c_1 \left(
g^{\beta_1\gamma_1}g^{\beta_2\gamma_2}\cdots g^{\beta_j\gamma_j}
    p_1^{\gamma_{j+1}}p_1^{\gamma_{j+2}}\cdots p_1^{\gamma_{j^{'}}}
\right)\\
& & + c_2 \left(
g^{\beta_1\gamma_1}g^{\beta_2\gamma_2}\cdots g^{\beta_{j-1}\gamma_{j-1}}
    p_1^{\beta_j}p_1^{\gamma_j}p_1^{\gamma_{j+1}}\cdots p_1^{\gamma_{j^{'}}}
\right)\\
& & +\cdots \\
& & + c_j \left(
g^{\beta_1\gamma_1}p_1^{\beta_2}p_1^{\beta_3}\cdots 
	p_1^{\gamma_2}p_1^{\gamma_3}\cdots p_1^{\gamma_j^{'}}
\right)\\
& &  + c_{j+1} \left(
p_1^{\beta_1}p_1^{\beta_2}\cdots p_1^{\beta_j}
	p_1^{\gamma_1}p_1^{\gamma_2}\cdots p_1^{\gamma_j^{'}}
\right),
\end{array}
\end{equation}

\begin{equation}
\begin{array}{rcrl}
\Gamma_- & = & Q^{\beta_1\gamma_1} & \left\{
c_{j+2}\left(
g^{\beta_2\gamma_2}g^{\beta_3\gamma_3}\cdots g^{\beta_j\gamma_j}
  p_1^{\gamma_{j+1}}p_1^{\gamma_{j+2}}\cdots p_1^{\gamma_j^{'}}
\right) \right. \\
& & & + c_{j+3} \left(
g^{\beta_2\gamma_2}g^{\beta_3\gamma_3}\cdots g^{\beta_{j-1}\gamma_{j-1}}
  p_1^{\beta_j}p_1^{\gamma_{j}}p_1^{\gamma_{j+1}}\cdots p_1^{\gamma_j^{'}}
\right)\\
& & & + \cdots \\
& & & + c_{2j}\left(
g^{\beta_2\gamma_2}p_1^{\beta_3}p_1^{\beta_4}\cdots p_1^{\beta_j}
  p_1^{\gamma_3}p_1^{\gamma_4}\cdots p_1^{\gamma_{j^{'}}}
\right)\\
& & & \left.
+ c_{2j+1}\left(
p_1^{\beta_2}p_1^{\beta_3}\cdots p_1^{\beta_{j}}
  p_1^{\gamma_2}p_1^{\gamma_3}\cdots p_1^{\gamma_{j^{'}}}
\right)
\right\};
\end{array}
\end{equation}

\item $(J_1, J_2, J_3) = (1, j, j)$, with $j \leq 1$

\begin{equation}
\begin{array}{rcrl}
\Gamma_+ & = & g^{\alpha_1\gamma_1} & \left\{
c_1\left(
g^{\beta_1\gamma_2}g^{\beta_2\gamma_3}\cdots g^{\beta_{j-1}\gamma_j}
    p_1^{\beta_j}
\right)\right.\\
& & & + c_2\left(
g^{\beta_1\gamma_2}g^{\beta_2\gamma_3}\cdots g^{\beta_{j-2}\gamma_{j-1}}
    p_1^{\beta_{j-1}}p_1^{\beta_j}p_1^{\gamma_j}
\right)\\
& & & + \cdots \\
& & & \left. + c_j \left(
p_1^{\beta_1}p_1^{\beta_2}\cdots p_1^{\beta_j}
    p_1^{\gamma_2}p_1^{\gamma_3}\cdots p_1^{\gamma_j}
\right)
\right\}\\
& & + p_2^{\alpha_1} & \left\{
c_{j+1}\left(
g^{\beta_1\gamma_1}g^{\beta_2\gamma_2}\cdots g^{\beta_j\gamma_j}
\right)\right. \\
& & & + c_{j+2}\left(
g^{\beta_1\gamma_1}g^{\beta_2\gamma_2}\cdots g^{\beta_{j-1}\gamma_{j-1}}
    p_1^{\beta_j}p_1^{\gamma_j}
\right)\\
& & & +\cdots \\
& & & \left. + c_{2j+1}\left(
p_1^{\beta_1}p_1^{\beta_2}\cdots p_1^{\beta_j}
    p_1^{\gamma_1}p_1^{\gamma_2}\cdots p_1^{\gamma_j}
\right)
\right\} \\
& & + g^{\alpha_1\beta_1} & \left\{
c_{2j+2}\left(
g^{\beta_2\gamma_1}g^{\beta_3\gamma_2}\cdots g^{\beta_j\gamma_{j-1}}
    p_1^{\gamma_j}
\right) \right.\\
& &  & + c_{2j+3}\left(
g^{\beta_2\gamma_1}g^{\beta_3\gamma_2}\cdots g^{\beta_{j-1}\gamma_{j-2}}
  p_1^{\beta_j}p_1^{\gamma_{j-1}}p_1^{\gamma_j}
\right)\\
& & & +\cdots \\
& & & \left. + c_{3j+1}\left(
p_1^{\beta_2}p_1^{\beta_3}\cdots p_1^{\beta_j}
    p_1^{\gamma_1}p_1^{\gamma_2}\cdots p_1^{\gamma_j}
\right)\right\},
\end{array}
\end{equation}

\begin{equation}
\begin{array}{rcrl}
\Gamma_- & = & A_1^{\alpha_1\beta_1\gamma_1} & \left\{
c_{3j+2}\left(
g^{\beta_2\gamma_2}g^{\beta_3\gamma_3}\cdots g^{\beta_j\gamma_j}
\right)\right.\\
& & & + c_{3j+3}\left(
g^{\beta_2\gamma_2}g^{\beta_3\gamma_3}\cdots g^{\beta_j\gamma_j}
\right)\\
& & & + \cdots \\
& & & \left. + c_{4j+1} \left(
p_1^{\beta_2}p_1^{\beta_2}\cdots p_1^{\beta_j}
    p_1^{\gamma_2}p_1^{\gamma_3}\cdots p_1^{\gamma_j}
\right)\right\}\\
& & + A_2^{\alpha_1\beta_1\gamma_1} & \left\{
c_{4j+2}\left(
g^{\beta_2\gamma_2}g^{\beta_3\gamma_3}\cdots g^{\beta_j\gamma_j}
\right)\right.\\
& & & + c_{4j+3}\left(
g^{\beta_2\gamma_2}g^{\beta_3\gamma_3}\cdots g^{\beta_j\gamma_j}
\right)\\
& & & + \cdots \\
& & & \left. + c_{5j+1} \left(
p_1^{\beta_2}p_1^{\beta_2}\cdots p_1^{\beta_j}
    p_1^{\gamma_2}p_1^{\gamma_3}\cdots p_1^{\gamma_j}
\right)\right\}\\
& & + Q^{\alpha_1\beta_1} & \left\{
c_{5j+2}\left(
g^{\beta_2\gamma_1}g^{\beta_3\gamma_2}\cdots g^{\beta_j\gamma_{j-1}}
    p_1^{\gamma_j}
\right)\right. \\
& & & + c_{5j+3}\left(
g^{\beta_2\gamma_1}g^{\beta_3\gamma_2}\cdots g^{\beta_{j-1}\gamma_{j-2}}
    p_1^{\beta_j}p_1^{\gamma_{j-1}}p_1^{\gamma_j}
\right)\\
& & & + \cdots \\
& & & \left. + c_{6j+1}\left(
p_1^{\beta_2}p_1^{\beta_3}\cdots p_1^{\beta_j}
    p_1^{\gamma_1}p_1^{\gamma_2}\cdots p_1^{\gamma_j}
\right)\right\};
\end{array}
\end{equation}

\item $(J_1, J_2, J_3) = (1, j, j^{'})$, with $1 \leq j < j^{'}$

\begin{equation}
\begin{array}{rcrl}
\Gamma_+ & = & g^{\alpha_1\gamma_1} & \left\{
c_1 \left(
g^{\beta_1\gamma_2}g^{\beta_2\gamma_3}\cdots g^{\beta_j\gamma_{j+1}}
    p_1^{\gamma_{j+2}}p_1^{\gamma_{j+3}}\cdots p_1^{\gamma_{j^{'}}}
\right) \right. \\
& & & +c_2 \left(
g^{\beta_1\gamma_2}g^{\beta_2\gamma_3}\cdots g^{\beta_{j-1}\gamma_j}
    p_1^{\beta_j}p_1^{\gamma_{j+1}}p_1^{\gamma_{j+2}}\cdots p_1^{\gamma_{j^{'}}}
\right)\\
& & & + \cdots \\
& & & +  \left. c_{j+1}\left(
p_1^{\beta_1}p_1^{\beta_2}\cdots p_1^{\beta_j}
    p_1^{\gamma_2}p_1^{\gamma_3}\cdots p_1^{\gamma_j^{'}}
\right) \right\} \\
& & + p_2^{\alpha_1} & \left\{
c_{j+1}\left(
g^{\beta_1\gamma_1}g^{\beta_2\gamma_2}\cdots g^{\beta_j\gamma_j}
    p_1^{\gamma_{j+1}}p_1^{\gamma_{j+2}}\cdots p_1^{\gamma_{j^{'}}}
\right) \right. \\
& & & + c_{j+2} \left(
g^{\beta_1\gamma_1}g^{\beta_2\gamma_2}\cdots g^{\beta_{j-1}\gamma_{j-1}}
    p_1^{\beta_j}p_1^{\gamma_{j}}p_1^{\gamma_{j+1}}\cdots p_1^{\gamma_{j^{'}}}
\right)\\
& & & +\cdots \\
& & & + \left. c_{2j+2}\left(
p_1^{\beta_1}p_1^{\beta_2}\cdots p_1^{\beta_j}
    p_1^{\gamma_1}p_1^{\gamma_2}\cdots p_1^{\gamma_j^{'}}
\right)
\right\}\\
& & + g^{\alpha_1\beta_1} & \left\{
c_{2j+3}\left(
g^{\beta_2\gamma_1}g^{\beta_3\gamma_2}\cdots g^{\beta_j\gamma_{j-1}}
    p_1^{\gamma_j}p_1^{\gamma_{j+1}}\cdots p_1^{\gamma_{j^{'}}}
\right) \right. \\
& & & + c_{2j+4}\left(
g^{\beta_2\gamma_1}g^{\beta_3\gamma_2}\cdots g^{\beta_{j-1}\gamma_{j-2}}
    p_1^{\beta_j}p_1^{\gamma_{j-1}}p_1^{\gamma_j}\cdots p_1^{\gamma_{j^{'}}}
\right)\\
& & & +\cdots \\
& & & \left. + c_{3j+2}\left(
p_1^{\beta_2}p_1^{\beta_3}\cdots p_1^{\beta_j}
    p_1^{\gamma_1}p_1^{\gamma_2}\cdots p_1^{\gamma_{j^{'}}}
\right) \right\} ,
\end{array}
\end{equation}

\begin{equation}
\begin{array}{rcrl}
\Gamma_- & = & A_1^{\alpha_1\beta_1\gamma_1} & \left\{
c_{3j+3}\left(
g^{\beta_2\gamma_2}g^{\beta_3\gamma_3}\cdots g^{\beta_j\gamma_j}
    p_1^{\gamma_{j+1}}p_1^{\gamma_{j+2}}\cdots p_1^{\gamma_{j^{'}}}
\right) \right.\\
& & & + c_{3j+4}\left(
g^{\beta_2\gamma_2}g^{\beta_3\gamma_3}\cdots g^{\beta_{j-1}\gamma_{j-1}}
    p_1^{\beta_j}p_1^{\gamma_j}p_1^{\gamma_{j+1}}\cdots p_1^{\gamma_{j^{'}}}
\right)\\
& & & +\cdots \\
& & & \left. + c_{4j+2}\left(
p_1^{\beta_2}p_1^{\beta_3}\cdots p_1^{\beta_j}
    p_1^{\gamma_2}p_1^{\gamma_3}\cdots p_1^{\gamma_j^{'}}
\right) \right\}\\
& & + A_2^{\alpha_1\beta_1\gamma_1} & \left\{
c_{4j+3}\left(
g^{\beta_2\gamma_2}g^{\beta_3\gamma_3}\cdots g^{\beta_j\gamma_j}
    p_1^{\gamma_{j+1}}p_1^{\gamma_{j+2}}\cdots p_1^{\gamma_{j^{'}}}
\right) \right.\\
& & & + c_{4j+4}\left(
g^{\beta_2\gamma_2}g^{\beta_3\gamma_3}\cdots g^{\beta_{j-1}\gamma_{j-1}}
    p_1^{\beta_j}p_1^{\gamma_j}p_1^{\gamma_{j+1}}\cdots p_1^{\gamma_{j^{'}}}
\right)\\
& & & +\cdots \\
& & & \left. + c_{5j+2}\left(
p_1^{\beta_2}p_1^{\beta_3}\cdots p_1^{\beta_j}
    p_1^{\gamma_2}p_1^{\gamma_3}\cdots p_1^{\gamma_j^{'}}
\right) \right\}\\
& & + Q^{\alpha_1\gamma_1} & c_{5j+3}\left(
g^{\beta_1\gamma_2}g^{\beta_2\gamma_3}\cdots g^{\beta_j\gamma_{j+1}}
    p_1^{\gamma_{j+2}}p_1^{\gamma_{j+3}}\cdots p_1^{\gamma_{j^{'}}}
\right)\\
& & + Q^{\alpha_1\beta_1} & \left\{
c_{5j+4}\left(
g^{\beta_2\gamma_1}g^{\beta_3\gamma_2}\cdots g^{\beta_j\gamma_{j-1}}
    p_1^{\gamma_j}p_1^{\gamma_{j+1}}\cdots p_1^{\gamma_{j^{'}}}
\right)
\right. \\
& & & + c_{5j+5}\left(
g^{\beta_2\gamma_1}g^{\beta_3\gamma_2}\cdots g^{\beta_{j-1}\gamma_{j-2}}
    p_1^{\beta_j}p_1^{\gamma_{j-1}}p_1^{\gamma_j}\cdots p_1^{\gamma_{j^{'}}}
\right) \\
& & & \left. + c_{6j+3}\left(
p_1^{\beta_2}p_1^{\beta_3}\cdots p_1^{\beta_j}
    p_1^{\gamma_1}p_1^{\gamma_2}\cdots p_1^{\gamma_{j^{'}}}
\right)
\right\};
\end{array}
\end{equation}

\item $(J_1, J_2, J_3) = (2, 2, 2)$

\begin{equation}
\begin{array}{rcl}
\Gamma_+ & = &\;\;\; c_1 g^{\alpha_1\beta_1}g^{\alpha_2\beta_2} 
  p_1^{\gamma_1}p_1^{\gamma_2} \\
& & +c_2 g^{\alpha_1\beta_1}g^{\alpha_2\gamma_1} 
  g^{\beta_2\gamma_2}\\
& & +c_3 g^{\alpha_1\beta_1}g^{\alpha_2\gamma_1}
  p_1^{\beta_2}p_1^{\gamma_2} \\
& & +c_4 g^{\alpha_1\beta_1}p_2^{\alpha_2}
  g^{\beta_2\gamma_1}p_1^{\gamma_2}\\
& & +c_5 g^{\alpha_1\beta_1}p_2^{\alpha_2}
  p_1^{\beta_2}p_1^{\gamma_1}p_1^{\gamma_2} \\
& & +c_6 g^{\alpha_1\gamma_1}g^{\alpha_2\gamma_2}
  p_1^{\beta_1}p_1^{\beta_2}\\
& &  +c_7 g^{\alpha_1\gamma_1}p_2^{\alpha_2}
  g^{\beta_1\gamma_2}p_1^{\beta_2}\\
& & +c_8 g^{\alpha_1\gamma_1}p_2^{\alpha_2}
  p_1^{\beta_1}p_1^{\beta_2}p_1^{\gamma_2}\\
& & + c_9 p_2^{\alpha_1}p_2^{\alpha_2}
  g^{\beta_1\gamma_1}g^{\beta_2\gamma_2}\\
& & + c_{10} p_2^{\alpha_1}p_2^{\alpha_2}
  p_1^{\beta_1}p_1^{\beta_2}p_1^{\gamma_1}p_1^{\gamma_2},
\end{array}
\end{equation}

\begin{equation}
\begin{array}{rcl}
\Gamma_- & = 
& \;\;\; c_{11} g^{\alpha_1\beta_1}A_1^{\alpha_2\beta_2\gamma_1}  
  p_1^{\gamma_2} \\
& & +c_{12} g^{\alpha_1\beta_1}A_2^{\alpha_2\beta_2\gamma_1}  
  p_1^{\gamma_2}\\
& & + c_{13} g^{\alpha_1\beta_1}Q^{\alpha_2\gamma_1}
  g^{\beta_2\gamma_2}\\
& & +c_{14} g^{\alpha_1\beta_1}Q^{\alpha_2\beta_2}
  p_1^{\gamma_1}p_1^{\gamma_2}\\
& & +c_{15} A_1^{\alpha_1\beta_1\gamma_1}g^{\alpha_2\gamma_2}
  p_1^{\beta_2}\\
& & +c_{16} A_1^{\alpha_1\beta_1\gamma_1}p_2^{\alpha_2}  
  g^{\beta_2\gamma_2}\\
& & +c_{17} A_1^{\alpha_1\beta_1\gamma_1}p_2^{\alpha_2}
  p_1^{\beta_2}p_1^{\gamma_2}\\
& & + c_{18} A_2^{\alpha_1\beta_1\gamma_1}p_2^{\alpha_2}
  g^{\beta_2\gamma_2}\\
& & + c_{19} Q^{\alpha_1\beta_1}p_2^{\alpha_2} 
  p_1^{\beta_2}p_1^{\gamma_1}p_1^{\gamma_2};
\end{array}
\end{equation}

\item $(J_1, J_2, J_3) = (2, j, j)$, with $j \geq 3$

\begin{equation}
\begin{array}{rcrl}
\Gamma_+ & = & g^{\alpha_1\beta_1}g^{\alpha_2\beta_2} & \left\{
c_1\left(
g^{\beta_3\gamma_1}g^{\beta_4\gamma_2}\cdots g^{\beta_j\gamma_{j-2}}
  p_1^{\gamma_{j-1}}p_1^{\gamma_j}
\right)\right. \\
& & & +c_2\left(
g^{\beta_3\gamma_1}g^{\beta_4\gamma_2}\cdots g^{\beta_{j-1}\gamma_{j-3}}
  p_1^{\beta_j}p_1^{\gamma_{j-2}}p_1^{\gamma_{j-1}}
	p_1^{\gamma_j}
\right) \\
& & & +\cdots \\
& & & \left. + c_{j-1} \left(
p_1^{\beta_3}p_1^{\beta_4}\cdots p_1^{\beta_j}
  p_1^{\gamma_1}p_1^{\gamma_2}\cdots p_1^{\gamma_j}
\right) \right\} \\
& & + g^{\alpha_1\beta_1}g^{\alpha_2\gamma_1} & \left\{
c_j\left(
g^{\beta_2\gamma_2}g^{\beta_3\gamma_3}\cdots g^{\beta_j\gamma_j}
\right) \right. \\
& & & +c_{j+1}\left(
g^{\beta_2\gamma_2}g^{\beta_3\gamma_3}\cdots g^{\beta_{j-1}\gamma_{j-1}}
  p_1^{\beta_j}p_1^{\gamma_j}
\right) \\
& & & +\cdots \\
& & & \left. +c_{2j-1}\left(
p_1^{\beta_2}p_1^{\beta_3}\cdots p_1^{\beta_j}
  p_1^{\gamma_2}p_1^{\gamma_3}\cdots p_1^{\gamma_j}
\right) \right\} \\
\end{array}
\end{equation}

\begin{equation}
\begin{array}{rcrl}
\Gamma_- & = 
& g^{\alpha_1\beta_1}g^{\alpha_2\beta_2}Q^{\beta_3\gamma_1} & \left\{
c_{5j+1}\left(
g^{\beta_4\gamma_2}g^{\beta_5\gamma_3}\cdots g^{\beta_j\gamma_{j-2}}
  p_1^{\gamma_{j-1}}p_1^{\gamma_j}
\right)\right. \\
& & & +c_{5j+2}\left(
g^{\beta_4\gamma_2}g^{\beta_5\gamma_3}\cdots g^{\beta_{j-1}\gamma_{j-3}}
  p_1^{\beta_j}p_1^{\gamma_{j-2}}p_1^{\gamma_{j-1}}p_1^{\gamma_j}
\right) \\
& & & +\cdots \\
& & & \left. +c_{6j-2}\left(
p_1^{\beta_4}p_1^{\beta_5}\cdots p_1^{\beta_j}
  p_1^{\gamma_2}p_1^{\gamma_3}\cdots p_1^{\gamma_j}
\right)\right\}\\
& & +g^{\alpha_1\beta_1}A_1^{\alpha_2\beta_2\gamma_1} & \left\{
c_{6j-1}\left(
g^{\beta_3\gamma_2}g^{\beta_4\gamma_3}\cdots g^{\beta_j\gamma_{j-1}}
  p_1^{\gamma_j}
\right)\right.\\
& & & +c_{6j}\left(
g^{\beta_3\gamma_2}g^{\beta_4\gamma_3}\cdots g^{\beta_{j-1}\gamma_{j-2}}
  p_1^{\beta_j}p_1^{\gamma_{j-1}}p_1^{\gamma_j}
\right)\\
& & & +\cdots \\
& & & \left. +c_{7j-3}\left(
p_1^{\beta_3}p_1^{\beta_4}\cdots p_1^{\beta_j}
  p_1^{\gamma_2}p_1^{\gamma_3}\cdots p_1^{\gamma_j}
\right)\right\}\\
\end{array}
\end{equation}

\item $(J_1, J_2, J_3) = (2, 2, 3)$

\begin{equation}
\begin{array}{rcl}
\Gamma_+ & = &\;\;\; c_1 g^{\alpha_1\beta_1}g^{\alpha_2\beta_2} 
  p_1^{\gamma_1}p_1^{\gamma_2}p_1^{\gamma_3} \\
& & +c_2 g^{\alpha_1\beta_1}g^{\alpha_2\gamma_1} 
  g^{\beta_2\gamma_2}p_1^{\gamma_3} \\
& & +c_3 g^{\alpha_1\beta_1}g^{\alpha_2\gamma_1}
  p_1^{\beta_2}p_1^{\gamma_2}p_1^{\gamma_3}\\
& & +c_4 g^{\alpha_1\beta_1}p_2^{\alpha_2}
  g^{\beta_2\gamma_1}p_1^{\gamma_2}p_1^{\gamma_3}\\
& & +c_5 g^{\alpha_1\beta_1}p_2^{\alpha_2}
  p_1^{\beta_2}p_1^{\gamma_1}p_1^{\gamma_2}p_1^{\gamma_3}\\
& & +c_6 g^{\alpha_1\gamma_1}g^{\alpha_2\gamma_2}
  g^{\beta_1\gamma_3}p_1^{\beta_2}\\
& & +c_7 g^{\alpha_1\gamma_1}g^{\alpha_2\gamma_2}
  p_1^{\beta_1}p_1^{\beta_2}p_1^{\gamma_3}\\
& & +c_8 g^{\alpha_1\gamma_1}p_2^{\alpha_2} 
  g^{\beta_1\gamma_2}g^{\beta_2\gamma_3}\\
& &  +c_9 g^{\alpha_1\gamma_1}p_2^{\alpha_2}
  g^{\beta_1\gamma_2}p_1^{\beta_2}p_1^{\gamma_3}\\
& & +c_{10} g^{\alpha_1\gamma_1}p_2^{\alpha_2}
  p_1^{\beta_1}p_1^{\beta_2}p_1^{\gamma_2}p_1^{\gamma_3}\\
& & + c_{11} p_2^{\alpha_1}p_2^{\alpha_2}
  g^{\beta_1\gamma_1}g^{\beta_2\gamma_2}p_1^{\gamma_3}\\
& & + c_{12} p_2^{\alpha_1}p_2^{\alpha_2}
  p_1^{\beta_1}p_1^{\beta_2}
	p_1^{\gamma_1}p_1^{\gamma_2}p_1^{\gamma_3},
\end{array}
\end{equation}

\begin{equation}
\begin{array}{rcl}
\Gamma_- & = 
& \;\;\; c_{13} g^{\alpha_1\beta_1}A_1^{\alpha_2\beta_2\gamma_1}  
  p_1^{\gamma_2}p_1^{\gamma_3}\\
& & +c_{14} g^{\alpha_1\beta_1}A_2^{\alpha_2\beta_2\gamma_1}  
  p_1^{\gamma_2}p_1^{\gamma_3}\\
& & + c_{15} g^{\alpha_1\beta_1}Q^{\alpha_2\gamma_1}
  g^{\beta_2\gamma_2}p^{\gamma_3}\\
& & +c_{16} g^{\alpha_1\beta_1}Q^{\alpha_2\beta_2}
  p_1^{\gamma_1}p_1^{\gamma_2}p_1^{\gamma_3}\\
& & +c_{17} A_1^{\alpha_1\beta_1\gamma_1}g^{\alpha_2\gamma_2}  
  g^{\beta_2\gamma_3}\\
& & +c_{18} A_1^{\alpha_1\beta_1\gamma_1}g^{\alpha_2\gamma_2}
  p_1^{\beta_2}p_1^{\gamma_3}\\
& & +c_{19} A_1^{\alpha_1\beta_1\gamma_1}p_2^{\alpha_2}  
  g^{\beta_2\gamma_2}p_1^{\gamma_3}\\
& & +c_{20} A_1^{\alpha_1\beta_1\gamma_1}p_2^{\alpha_2}
  p_1^{\beta_2}p_1^{\gamma_2}p_1^{\gamma_3}\\
& & + c_{21} A_2^{\alpha_1\beta_1\gamma_1}g^{\alpha_2\gamma_2}
  g^{\beta_2\gamma_3}\\
& & + c_{22} A_2^{\alpha_1\beta_1\gamma_1}p_2^{\alpha_2}
  g^{\beta_2\gamma_2}p_1^{\gamma_3}\\
& & + c_{23} Q^{\alpha_1\beta_1}p_2^{\alpha_2}
  p_1^{\beta_2}p_1^{\gamma_1}p_1^{\gamma_2}p_1^{\gamma_3};
\end{array}
\end{equation}

\item $(J_1, J_2, J_3) = (2, j, j+1)$, with $j \geq 3$

\begin{equation}
\begin{array}{rcrl}
\Gamma_+ & = & g^{\alpha_1\beta_1}g^{\alpha_2\beta_2} & \left\{
c_1\left(
g^{\beta_3\gamma_1}g^{\beta_4\gamma_2}\cdots g^{\beta_j\gamma_{j-2}}
  p_1^{\gamma_{j-1}}p_1^{\gamma_j}p_1^{\gamma_{j+1}}
\right)\right. \\
& & & +c_2\left(
g^{\beta_3\gamma_1}g^{\beta_4\gamma_2}\cdots g^{\beta_{j-1}\gamma_{j-3}}
  p_1^{\beta_j}p_1^{\gamma_{j-2}}p_1^{\gamma_{j-1}}
	p_1^{\gamma_j}p_1^{\gamma_{j+1}}
\right) \\
& & & +\cdots \\
& & & \left. + c_{j-1} \left(
p_1^{\beta_3}p_1^{\beta_4}\cdots p_1^{\beta_j}
  p_1^{\gamma_1}p_1^{\gamma_2}\cdots p_1^{\gamma_{j+1}}
\right) \right\} \\
& & + g^{\alpha_1\beta_1}g^{\alpha_2\gamma_1} & \left\{
c_j\left(
g^{\beta_2\gamma_2}g^{\beta_3\gamma_3}\cdots g^{\beta_j\gamma_j}
  p_1^{\gamma_{j+1}}
\right) \right. \\
& & & +c_{j+1}\left(
g^{\beta_2\gamma_2}g^{\beta_3\gamma_3}\cdots g^{\beta_{j-1}\gamma_{j-1}}
  p_1^{\beta_j}p_1^{\gamma_j}p_1^{\gamma_{j+1}}
\right) \\
& & & +\cdots \\
& & & \left. +c_{2j-1}\left(
p_1^{\beta_2}p_1^{\beta_3}\cdots p_1^{\beta_j}
  p_1^{\gamma_2}p_1^{\gamma_3}\cdots p_1^{\gamma_j}p_1^{\gamma_{j+1}}
\right) \right\} \\
\end{array}
\end{equation}

\begin{equation}
\begin{array}{rcrl}
\Gamma_- & = 
& g^{\alpha_1\beta_1}g^{\alpha_2\beta_2}Q^{\beta_3\gamma_1} & \left\{
c_{5j+3}\left(
g^{\beta_4\gamma_2}g^{\beta_5\gamma_3}\cdots g^{\beta_j\gamma_{j-2}}
  p_1^{\gamma_{j-1}}p_1^{\gamma_j}p_1^{\gamma_{j+1}}
\right)\right. \\
& & & +c_{5j+4}\left(
g^{\beta_4\gamma_2}g^{\beta_5\gamma_3}\cdots g^{\beta_{j-1}\gamma_{j-3}}
  p_1^{\beta_j}p_1^{\gamma_{j-2}}p_1^{\gamma_{j-1}}p_1^{\gamma_j}
	p_1^{\gamma_{j+1}}
\right) \\
& & & +\cdots \\
& & & \left. +c_{6j}\left(
p_1^{\beta_4}p_1^{\beta_5}\cdots p_1^{\beta_j}
  p_1^{\gamma_2}p_1^{\gamma_3}\cdots p_1^{\gamma_{j+1}}
\right)\right\}\\
& & +g^{\alpha_1\beta_1}A_1^{\alpha_2\beta_2\gamma_1} & \left\{
c_{6j+1}\left(
g^{\beta_3\gamma_2}g^{\beta_4\gamma_3}\cdots g^{\beta_j\gamma_{j-1}}
  p_1^{\gamma_j}p_1^{\gamma_{j+1}}
\right)\right.\\
& & & +c_{6j+2}\left(
g^{\beta_3\gamma_2}g^{\beta_4\gamma_3}\cdots g^{\beta_{j-1}\gamma_{j-2}}
  p_1^{\beta_j}p_1^{\gamma_{j-1}}p_1^{\gamma_j}p_1^{\gamma_{j+1}}
\right)\\
& & & +\cdots \\
& & & \left. +c_{7j-1}\left(
p_1^{\beta_3}p_1^{\beta_4}\cdots p_1^{\beta_j}
  p_1^{\gamma_2}p_1^{\gamma_3}\cdots p_1^{\gamma_{j+1}}
\right)\right\}\\
\end{array}
\end{equation}

\item $(J_1, J_2, J_3) = (2, 2, j)$, with $j\geq 4$

\begin{equation}
\begin{array}{rcl}
\Gamma_+ & = &\;\;\; c_1 g^{\alpha_1\beta_1}g^{\alpha_2\beta_2} 
\left(
  p_1^{\gamma_1}p_1^{\gamma_2}\cdots p_1^{\gamma_j}
\right) \\
& & +c_2 g^{\alpha_1\beta_1}g^{\alpha_2\gamma_1} 
\left(
g^{\beta_2\gamma_2}
  p_1^{\gamma_3}p_1^{\gamma_4}\cdots p_1^{\gamma_j}
\right) \\
& & +c_3 g^{\alpha_1\beta_1}g^{\alpha_2\gamma_1}
\left(
  p_1^{\beta_2}p_1^{\gamma_2}p_1^{\gamma_3}\cdots p_1^{\gamma_j}
\right) \\
& & +c_4 g^{\alpha_1\beta_1}p_2^{\alpha_2}
\left(
g^{\beta_2\gamma_1}
  p_1^{\gamma_2}p_1^{\gamma_3}\cdots p_1^{\gamma_j}
\right) \\
& & +c_5 g^{\alpha_1\beta_1}p_2^{\alpha_2}
\left(
p_1^{\beta_2}
  p_1^{\gamma_1}p_1^{\gamma_2}\cdots p_1^{\gamma_j}
\right) \\
& & +c_6 g^{\alpha_1\gamma_1}g^{\alpha_2\gamma_2} 
\left(
g^{\beta_1\gamma_3}g^{\beta_2\gamma_4}
  p_1^{\gamma_5}p_1^{\gamma_6}\cdots p_1^{\gamma_j}
\right) \\
& & +c_7 g^{\alpha_1\gamma_1}g^{\alpha_2\gamma_2}
\left(
g^{\beta_1\gamma_3}
  p_1^{\beta_2}p_1^{\gamma_4}p_1^{\gamma_5}\cdots p_1^{\gamma_j}
\right) \\
& & +c_8 g^{\alpha_1\gamma_1}g^{\alpha_2\gamma_2}
\left(
p_1^{\beta_1}p_1^{\beta_2}
  p_1^{\gamma_3}p_1^{\gamma_4}\cdots p_1^{\gamma_j}
\right) \\
& & +c_9 g^{\alpha_1\gamma_1}p_2^{\alpha_2} 
\left(
g^{\beta_1\gamma_2}g^{\beta_2\gamma_3}
  p_1^{\gamma_4}p_1^{\gamma_5}\cdots p_1^{\gamma_j}
\right) \\
& &  +c_{10} g^{\alpha_1\gamma_1}p_2^{\alpha_2}
\left(
g^{\beta_1\gamma_2}
  p_1^{\beta_2}p_1^{\gamma_3}p_1^{\gamma_4}\cdots p_1^{\gamma_j}
\right) \\
& & +c_{11} g^{\alpha_1\gamma_1}p_2^{\alpha_2}
\left(
p_1^{\beta_1}p_1^{\beta_2}
  p_1^{\gamma_2}p_1^{\gamma_3}\cdots p_1^{\gamma_j}
\right) \\
& & + c_{12} p_2^{\alpha_1}p_2^{\alpha_2} \left(
g^{\beta_1\gamma_1}g^{\beta_2\gamma_2}
  p_1^{\gamma_3}p_1^{\gamma_4}\cdots p_1^{\gamma_j}
\right)\\
& & + c_{13} p_2^{\alpha_1}p_2^{\alpha_2} \left(
p_1^{\beta_1}p_1^{\beta_2}
  p_1^{\gamma_1}p_1^{\gamma_2}\cdots p_1^{\gamma_j}
\right),
\end{array}
\end{equation}

\begin{equation}
\begin{array}{rcl}
\Gamma_- & = 
& \;\;\; c_{14} g^{\alpha_1\beta_1}A_1^{\alpha_2\beta_2\gamma_1}  
\left(
p_1^{\gamma_2}p_1^{\gamma_3}\cdots p_1^{\gamma_j}
\right) \\
& & +c_{15} g^{\alpha_1\beta_1}A_2^{\alpha_2\beta_2\gamma_1}  
\left(
  p_1^{\gamma_2}p_1^{\gamma_3}\cdots p_1^{\gamma_j}
\right) \\
& & + c_{16} g^{\alpha_1\beta_1}Q^{\alpha_2\gamma_1} \left(
g^{\beta_2\gamma_2}
  p^{\gamma_3}p^{\gamma_4}\cdots p^{\gamma_j}
\right)\\
& & +c_{17} g^{\alpha_1\beta_1}Q^{\alpha_2\beta_2} \left(
  p_1^{\gamma_1}p_1^{\gamma_2}\cdots p_1^{\gamma_j}
\right)\\
& & +c_{18} A_1^{\alpha_1\beta_1\gamma_1}g^{\alpha_2\gamma_2}  
\left(
g^{\beta_2\gamma_3}
  p_1^{\gamma_4}p_1^{\gamma_5}\cdots p_1^{\gamma_j}
\right) \\
& & +c_{19} A_1^{\alpha_1\beta_1\gamma_1}g^{\alpha_2\gamma_2}
\left(
p_1^{\beta_2}
  p_1^{\gamma_3}p_1^{\gamma_4}\cdots p_1^{\gamma_j}
\right)\\
& & +c_{20} A_1^{\alpha_1\beta_1\gamma_1}p_2^{\alpha_2}  
\left(
g^{\beta_2\gamma_2}
  p_1^{\gamma_3}p_1^{\gamma_4}\cdots p_1^{\gamma_j}
\right) \\
& & +c_{21} A_1^{\alpha_1\beta_1\gamma_1}p_2^{\alpha_2}
\left(
p_1^{\beta_2}
  p_1^{\gamma_2}p_1^{\gamma_3}\cdots p_1^{\gamma_j}
\right)\\
& & + c_{22} A_2^{\alpha_1\beta_1\gamma_1}g^{\alpha_2\gamma_2} \left(
g^{\beta_2\gamma_3}
  p_1^{\gamma_4}p_1^{\gamma_5}\cdots p_1^{\gamma_j}
\right)\\
& & + c_{23} A_2^{\alpha_1\beta_1\gamma_1}p_2^{\alpha_2} \left(
g^{\beta_2\gamma_2}
  p_1^{\gamma_3}p_1^{\gamma_4}\cdots p_1^{\gamma_j}
\right)\\
& & +c_{24} Q^{\alpha_1\gamma_1}g^{\alpha_2\gamma_2} \left(
g^{\beta_1\gamma_3}g^{\beta_2\gamma_4}
  p_1^{\gamma_5}p_1^{\gamma_6}\cdots p_1^{\gamma_j}
\right)\\
& & + c_{25} Q^{\alpha_1\beta_1}p_2^{\alpha_2} \left(
p_1^{\beta_2}
  p_1^{\gamma_1}p_1^{\gamma_2}\cdots p_1^{\gamma_j}
\right);
\end{array}
\end{equation}

\item $(J_1, J_2, J_3) = (2, j, j^{'})$, with $j \geq 3$ and $j^{'}\geq j+2$

\begin{equation}
\begin{array}{rcrl}
\Gamma_+ & = & g^{\alpha_1\beta_1}g^{\alpha_2\beta_2} & \left\{
c_1\left(
g^{\beta_3\gamma_1}g^{\beta_4\gamma_2}\cdots g^{\beta_j\gamma_{j-2}}
  p_1^{\gamma_{j-1}}p_1^{\gamma_j}\cdots p_1^{\gamma_{j^{'}}}
\right)\right. \\
& & & +c_2\left(
g^{\beta_3\gamma_1}g^{\beta_4\gamma_2}\cdots g^{\beta_{j-1}\gamma_{j-3}}
  p_1^{\beta_j}p_1^{\gamma_{j-2}}p_1^{\gamma_{j-1}}\cdots p_1^{\gamma_{j^{'}}}
\right)\\
& & & +\cdots \\
& & & \left. + c_{j-1} \left(
p_1^{\beta_3}p_1^{\beta_4}\cdots p_1^{\beta_j}
  p_1^{\gamma_1}p_1^{\gamma_2}\cdots p_1^{\gamma_{j^{'}}}
\right) \right\} \\
& & + g^{\alpha_1\beta_1}g^{\alpha_2\gamma_1} & \left\{
c_j\left(
g^{\beta_2\gamma_2}g^{\beta_3\gamma_3}\cdots g^{\beta_j\gamma_j}
  p_1^{\gamma_{j+1}}p_1^{\gamma_{j+2}}\cdots p_1^{\gamma_{j^{'}}}
\right) \right. \\
& & & +c_{j+1}\left(
g^{\beta_2\gamma_2}g^{\beta_3\gamma_3}\cdots g^{\beta_{j-1}\gamma_{j-1}}
  p_1^{\beta_j}p_1^{\gamma_j}p_1^{\gamma_{j+1}}\cdots p_1^{\gamma_{j^{'}}}
\right) \\
& & & +\cdots \\
& & & \left. +c_{2j-1}\left(
p_1^{\beta_2}p_1^{\beta_3}\cdots p_1^{\beta_j}
  p_1^{\gamma_2}p_1^{\gamma_3}\cdots p_1^{\gamma_{j^{'}}}
\right) \right\} \\
\end{array}
\end{equation}

\begin{equation}
\begin{array}{rcrl}
\Gamma_- & = 
& g^{\alpha_1\beta_1}g^{\alpha_2\beta_2}Q^{\beta_3\gamma_1} & \left\{
c_{5j+4}\left(
g^{\beta_4\gamma_2}g^{\beta_5\gamma_3}\cdots g^{\beta_j\gamma_{j-2}}
  p_1^{\gamma_{j-1}}p_1^{\gamma_j}\cdots p_1^{\gamma_{j^{'}}}
\right)\right. \\
& & & +c_{5j+5}\left(
g^{\beta_4\gamma_2}g^{\beta_5\gamma_3}\cdots g^{\beta_{j-1}\gamma_{j-3}}
  p_1^{\beta_j}p_1^{\gamma_{j-2}}p_1^{\gamma_{j-1}}\cdots p_1^{\gamma_{j^{'}}}
\right) \\
& & & +\cdots \\
& & & \left. +c_{6j+1}\left(
p_1^{\beta_4}p_1^{\beta_5}\cdots p_1^{\beta_j}
  p_1^{\gamma_2}p_1^{\gamma_3}\cdots p_1^{\gamma_{j^{'}}}
\right)\right\}\\
& & +g^{\alpha_1\beta_1}A_1^{\alpha_2\beta_2\gamma_1} & \left\{
c_{6j+2}\left(
g^{\beta_3\gamma_2}g^{\beta_4\gamma_3}\cdots g^{\beta_j\gamma_{j-1}}
  p_1^{\gamma_j}p_1^{\gamma_{j+1}}\cdots p_1^{\gamma_{j^{'}}}
\right)\right.\\
& & & +c_{6j+3}\left(
g^{\beta_3\gamma_2}g^{\beta_4\gamma_3}\cdots g^{\beta_{j-1}\gamma_{j-2}}
  p_1^{\beta_j}p_1^{\gamma_{j-1}}p_1^{\gamma_j}\cdots p_1^{\gamma_{j^{'}}}
\right)\\
& & & +\cdots \\
& & & \left. +c_{7j}\left(
p_1^{\beta_3}p_1^{\beta_4}\cdots p_1^{\beta_j}
  p_1^{\gamma_2}p_1^{\gamma_3}\cdots p_1^{\gamma_{j^{'}}}
\right)\right\}\\
\end{array}
\end{equation}

\end{itemize}
etc. One can give more effective vertexes for those cases in which all three 
spins $\geq 3$, but this will take too much space and these vertexes are 
seldom used. We drop the factor $\delta (p_1+p_2+p_3)$ in effective vertexes
and amplitudes through out this paper. Scalar coefficients $c_i$ are complex 
functions of the form
\begin{equation}
c_i = c_i(p_1^2, p_2^2, p_1\cdot p_2).
\end{equation}
They are just constants when all of the particles are on shell.  
These coefficients are also called invariant amplitudes, coupling
coefficients, structure functions or form factors in different references.

The propagator of an off shell particle does not satisfy 
Eq.~{(\ref{eq:rs1})} and Eq.~{(\ref{eq:rs3})}. In this case,
those terms proportional to
\begin{equation}
p_1^{\alpha_i},\; p_2^{\beta_i},\; (p_1+p_2)^{\gamma_i},\;
g^{\alpha_i\alpha_j},\; g^{\beta_i\beta_j},\; g^{\gamma_i\gamma_j}
\end{equation} 
will not vanish. However, such an off shell term always contributes a factor 
of $p_1^2-W_1^2$, $p_2^2-W_2^2$ or $p_3^2-W_3^2$ after contraction
with the propagator. This factor will eliminate the denominator
of the propagator, which makes the amplitude free of pole at the 
mass of the intermediate particle\footnote{See Ref.\cite{WeinbergQFT1} for 
arguments on polology. }. 
These contributions are not independent from background terms.
They can be absorbed into back ground amplitudes without introducing 
kinematic singularities. 
Off shell terms are not needed when back ground amplitudes are included. 
One can refer to Sec.~\ref{sec:bk} or Ref.~\cite{Zhua23pi} for examples.

\section{Symmetry under space reflection}\label{sec:sr}
Under space reflection operation ${\bf P}$, canonical quantum states 
transform as
\begin{equation}
{\bf P}|\roarrow{p}, j, m\rangle = \eta |-\roarrow{p}, j, m\rangle.
\end{equation} 
If parity is conserved
\begin{equation}
{\bf P}^\dagger {\bf S} {\bf P} = {\bf S},
\end{equation}
\begin{equation}\label{eq:pp}
\begin{array}{rl}
& \langle \roarrow{p}_1^{'},J_1^{'},m_1^{'};\roarrow{p}_2^{'},J_2^{'},m_2^{'};
\cdots|{\bf S}|\roarrow{p}_1,J_1,m_1;\roarrow{p}_2,J_2,m_2;\cdots\rangle \\
= & {\eta_1^{'}}^{\ast}{\eta_2^{'}}^{\ast}\cdots
\eta_1\eta_2\cdots\langle -\roarrow{p}_1^{'},J_1^{'},m_1^{'};
-\roarrow{p}_2^{'},J_2^{'},m_2^{'};\cdots
|{\bf S}|-\roarrow{p}_1,J_1,m_1;-\roarrow{p}_2,J_2,m_2;\cdots\rangle,
\end{array}
\end{equation}
where ${\bf S}$ is the S-matrix operator.

The space reflection matrix is defined as~\cite{ChungSF71}
\begin{equation}
\left( {P^\mu}_{\nu} \right) = {\rm diag}\{1, -1, -1, -1\},
\end{equation}
we have
\begin{eqnarray}
\bar{p}_i^\mu & \equiv & {P^\mu}_{\nu}p_i^\nu 
	= \left(E_i, -\roarrow{p}_i\right),\\
\bar{p}_i^{'\mu} & \equiv & {P^\mu}_{\nu}p_i^{'\nu}
	= \left(E^{'}_i, -\roarrow{p}^{'}_i\right),\\
\bar{g}^{\mu\nu} & \equiv & {P^\mu}_{\alpha} {P^\nu}_{\beta} g^{\alpha\beta}
	= g^{\mu\nu},\\
\bar{\varepsilon}^{\alpha\beta\gamma\delta} & \equiv &
	{P^\alpha}_{\alpha^{'}}{P^\beta}_{\beta^{'}}
	{P^\gamma}_{\gamma^{'}}{P^\delta}_{\delta^{'}}
	\varepsilon^{\alpha^{'}\beta^{'}\gamma^{'}\delta^{'}}
	= - \varepsilon^{\alpha\beta\gamma\delta},\\
\bar{e}^{\mu_1\mu_2\cdots\mu_j} & \equiv &
	{P^{\mu_1}}_{\nu_1}{P^{\mu_2}}_{\nu_2}\cdots{P^{\mu_j}}_{\nu_j}
	e^{\nu_1\nu_2\cdots \nu_j}.
\end{eqnarray}
Eq.~{(\ref{eq:pp})} reads
\begin{equation}
\begin{array}{rl}
& e^{\ast}_c(p_1^{'},J_1^{'},m_1^{'})e^{\ast}_c(p_2^{'},J_2^{'},m_2^{'})\cdots
e_c(p_1,J_1,m_1)e_c(p_2,J_2,m_2)\cdots\times\\
& \times\Gamma(p_1^{'},p_2^{'},\cdots,p_1,p_2,\cdots,g^{\mu\nu},
	\varepsilon^{\alpha\beta\gamma\delta})\\
= & {\eta_1^{'}}^{\ast}{\eta_2^{'}}^{\ast}\cdots\eta_1\eta_2\cdots
	e^{\ast}_c(\bar{p}_1^{'},J_1^{'},m_1^{'})
	e^{\ast}_c(\bar{p}_2^{'},J_2^{'},m_2^{'})\cdots
	e_c(\bar{p}_1,J_1,m_1)e_c(\bar{p}_2,J_2,m_2)\cdots\times\\
& \times\Gamma(\bar{p}_1^{'},\bar{p}_2^{'},\cdots,
		\bar{p}_1,\bar{p}_2,\cdots,g^{\mu\nu},
	\varepsilon^{\alpha\beta\gamma\delta}).
\end{array}
\end{equation}
From Eq.~{(\ref{eq:wfc})} one can see
\begin{equation}
e_c^\mu (\bar{p}, m) = - \bar{e}_c^\mu (p, m),
\end{equation}
and for spin-$j$ canonical wave functions,
\begin{equation}
e_c^{\mu_1\mu_2\cdots\mu_j} (\bar{p},j,m) = 
	(-1)^j \bar{e}_c^{\mu_1\mu_2\cdots\mu_j} (p,j,m).
\end{equation}
The requirement of parity conservation becomes
\begin{equation}
\begin{array}{rl}
& e^{\ast}_c(p_1^{'},J_1^{'},m_1^{'})e^{\ast}_c(p_2^{'},J_2^{'},m_2^{'})\cdots
e_c(p_1,J_1,m_1)e_c(p_2,J_2,m_2)\cdots\times\\
& \times\Gamma(p_1^{'},p_2^{'},\cdots,p_1,p_2,\cdots,g^{\mu\nu},
	\varepsilon^{\alpha\beta\gamma\delta})\\
= & {\eta_1^{'}}^{\ast}{\eta_2^{'}}^{\ast}\cdots\eta_1\eta_2\cdots
	(-1)^{J_1+J_2+\cdots-J_1^{'}-J_2^{'}-\cdots}
	e^{\ast}_c(p_1^{'},J_1^{'},m_1^{'})
	e^{\ast}_c(p_2^{'},J_2^{'},m_2^{'})\cdots
	e_c(p_1,J_1,m_1)e_c(p_2,J_2,m_2)\cdots\times\\
& \times\Gamma(p_1^{'},p_2^{'},\cdots,
		p_1,p_2,\cdots,\bar{g}^{\mu\nu},
	\bar{\varepsilon}^{\alpha\beta\gamma\delta})
\end{array}
\end{equation}
or
\begin{equation}
\begin{array}{rl}
&\Gamma(p_1^{'},p_2^{'},\cdots,p_1,p_2,\cdots,g^{\mu\nu},
	\varepsilon^{\alpha\beta\gamma\delta})\\
= & {\eta_1^{'}}^{\ast}{\eta_2^{'}}^{\ast}\cdots\eta_1\eta_2\cdots
	(-1)^{J_1+J_2+\cdots-J_1^{'}-J_2^{'}-\cdots} 
\Gamma(p_1^{'},p_2^{'},\cdots,p_1,p_2,\cdots,g^{\mu\nu},
	-\varepsilon^{\alpha\beta\gamma\delta}).
\end{array}
\end{equation}
That is, if space reflection parity is conserved, the effective vertexes
will consist of only tensors for the case
\begin{equation}
{\eta_1^{'}}^{\ast}{\eta_2^{'}}^{\ast}\cdots\eta_1\eta_2\cdots
	(-1)^{J_1+J_2+\cdots-J_1^{'}-J_2^{'}-\cdots} = 1,
\end{equation}
and  only pseudo-tensors if
\begin{equation}
{\eta_1^{'}}^{\ast}{\eta_2^{'}}^{\ast}\cdots\eta_1\eta_2\cdots
	(-1)^{J_1+J_2+\cdots-J_1^{'}-J_2^{'}-\cdots} = - 1.
\end{equation}
{\em Mixing of tensor and pseudo-tensor vertexes always means violation
of space reflection symmetry.}
One can also prove the same result using helicity wave 
functions~\cite{Zhua23pi}. 

We can see from Eq.~{(\ref{eq:00j-})} that,
if space reflection parity is conserved,
a spin-$j$ particle decaying into two scalar(or pseudo-scalar)
particles must has the parity of $(-1)^j$, while a particle decaying into
a scalar and a pseudo-scalar particle has the parity of $-(-1)^j$. 
Similarly, for spin-$0$ $\rightarrow$ spin-$j$ + spin-$0$,
the parity of the spin-$j$ particle will be $(-1)^j$ (or $-(-1)^j$)
if the two spin-$0$ particles have the same ( or opposite) parities.

\section{Boson symmetry}\label{sec:bs}
For two identical bosons, we have
\begin{equation}
\left[
a^\dagger( \roarrow{p}, j, \sigma ),
a^\dagger( \roarrow{p}^{'}, j, \sigma^{'} )
\right]
=0,
\end{equation}
\begin{equation}
\left|\cdots ;\roarrow{p},j,\sigma; \cdots; \roarrow{p}^{'},j,\sigma^{'};
	\cdots \right>
=
\left|\cdots ;\roarrow{p}^{'},j,\sigma^{'}; \cdots; \roarrow{p},j,\sigma;
	\cdots \right>.
\end{equation}
This demands\footnote{Strictly speaking, we should use Green functions to
derive properties of off shell effective vertexes.}
\begin{equation}\label{eq:bs}
\Gamma^{\cdots \mu_1\mu_2\cdots\mu_j\cdots
	\nu_1\nu_2\cdots\nu_j\cdots} =
\Gamma^{\cdots \nu_1\nu_2\cdots\nu_j\cdots
	\mu_1\mu_2\cdots\mu_j\cdots}.
\end{equation}

Suppose particle 1 and particle 2 are identical particles. We define
\begin{equation}
p=p_1+p_2,\;k=p_1-p_2,
\end{equation}
\begin{eqnarray}
A_+^{\alpha\beta\gamma} & = & k_\mu \varepsilon^{\mu\alpha\beta\gamma},\\
A_-^{\alpha\beta\gamma} & = & p_\mu \varepsilon^{\mu\alpha\beta\gamma}.
\end{eqnarray}

Let's first separate the 3-leg effective vertexes given in Sec.~\ref{sec:3v} 
into 1-2 symmetric parts $\Gamma_\pm^{(S)}$
and 1-2 anti-symmetric parts $\Gamma_\pm^{(A)}$:

\begin{itemize}

\item $(J_1, J_2, J_3) = (0, 0, j)$, with $j$ an even number

\begin{eqnarray}
\Gamma_+^{(S)} & = & c_1 (k^{\gamma_1}k^{\gamma_2}\cdots k^{\gamma_j}),\\
\Gamma_+^{(A)} & = & 0, \\
\Gamma_-^{(S,A)} & = & 0. 
\end{eqnarray}

\item $(J_1, J_2, J_3) = (0, 0, j)$, with $j$ an odd number

Similar to the previous case, with 
$\Gamma_\pm^{(S)}\longleftrightarrow \Gamma_\pm^{(A)}$:
\begin{eqnarray}
\Gamma_+^{(S)} & = & 0, \label{eq:00j.1}\\
\Gamma_+^{(A)} & = & c_1 (k^{\gamma_1}k^{\gamma_2}\cdots k^{\gamma_j}),\\
\Gamma_-^{(S,A)} & = & 0. \label{eq:00j.2}
\end{eqnarray}

\item $(J_1, J_2, J_3) = (j, j, 0)$, with $j \geq 1$

\begin{equation}
\begin{array}{rcl}
\Gamma_+^{(S)} & = &
\;\;\; c_1 \left( 
g^{\alpha_1\beta_1}g^{\alpha_2\beta_2}\cdots g^{\alpha_j\beta_j}
\right)\\
& & + c_2 \left(
g^{\alpha_1\beta_1}g^{\alpha_2\beta_2}\cdots g^{\alpha_{j-1}\beta_{j-1}}
p^{\alpha_j\beta_j}
\right)\\
& & + \cdots \\
& & + c_{j+1} \left(
p^{\alpha_1}p^{\alpha_2}\cdots p^{\alpha_j}
	p^{\beta_1}p^{\beta_2}\cdots p^{\beta_j}
\right),
\end{array}
\end{equation}

\begin{equation}
\Gamma_+^{(A)} = 0,
\end{equation}

\begin{equation}
\begin{array}{rcrl}
\Gamma_-^{(S)} & = & Q^{\alpha_1\beta_1} & \left\{
\; c_{j+2} \left(
g^{\alpha_2\beta_2}g^{\alpha_3\beta_3}\cdots g^{\alpha_j\beta_j}
\right)\right. \\
& & & +c_{j+3} \left(
g^{\alpha_2\beta_2}g^{\alpha_3\beta_3}\cdots g^{\alpha_{j-1}\beta_{j-1}}
	p^{\alpha_j}p^{\beta_j}
\right)\\
& & & +\cdots \\
& & & \left. +c_{2j+1} \left(
p^{\alpha_1}p^{\alpha_2}\cdots p^{\alpha_j}
	p^{\beta_1}p^{\beta_2}\cdots p^{\beta_j}
\right)\right\},
\end{array}
\end{equation}

\begin{equation}
\Gamma_-^{(A)} = 0.
\end{equation}

\item $(J_1, J_2, J_3) = (1, 1, j)$, with $j$ an even number and $j \geq 2$ 

\begin{equation}
\begin{array}{rcl}
\Gamma_+^{(S)} & = & \;\;\;c_1 
g^{\alpha_1\gamma_1}g^{\beta_1\gamma_2}
\left(
k^{\gamma_3}k^{\gamma_4}\cdots k^{\gamma_j}
\right)\\
& & +c_2 \left(
g^{\alpha_1\gamma_1}p^{\beta_1}
	-g^{\beta_1\gamma_1}p^{\alpha_1} \right)
\left( k^{\gamma_2}k^{\gamma_3}\cdots k^{\gamma_j} \right) \\
& & +c_3 g^{\alpha_1\beta_1} \left(
k^{\gamma_1}k^{\gamma_2}\cdots k^{\gamma_j} 
\right)\\
& & +c_4 p^{\alpha_1}p^{\beta_1} \left(
k^{\gamma_1}k^{\gamma_2}\cdots k^{\gamma_j}
\right),
\end{array}
\end{equation}

\begin{equation}
\Gamma_+^{(A)}  =  c_5 \left(
g^{\alpha_1\gamma_1}p^{\beta_1}
	+g^{\beta_1\gamma_1}p^{\alpha_1} \right)
\left( k^{\gamma_2}k^{\gamma_3}\cdots k^{\gamma_j} \right),
\end{equation}

\begin{equation}
\begin{array}{rcl}
\Gamma_-^{(S)} & = & 
\;\;\; c_6 A_-^{\alpha_1\beta_1\gamma_1}\left(
k^{\gamma_2}k^{\gamma_3}\cdots k^{\gamma_j}
\right)\\
& & +c_7 Q^{\alpha_1\beta_1}\left(
k^{\gamma_1}k^{\gamma_2}\cdots k^{\gamma_j}
\right),
\end{array}
\end{equation}

\begin{equation}
\begin{array}{rcl}
\Gamma_-^{(A)} & = & 
\;\;\; c_8 \left( Q^{\alpha_1\gamma_1}g^{\beta_1\gamma_2} +
  Q^{\beta_1\gamma_1}g^{\alpha_1\gamma_2} \right) \left(
k^{\gamma_3}k^{\gamma_4}\cdots k^{\gamma_j}
\right)\\
& & +c_9 A_+^{\alpha_1\beta_1\gamma_1}\left(
k^{\gamma_2}k^{\gamma_3}\cdots k^{\gamma_j}
\right).
\end{array}
\end{equation}

\item $(J_1, J_2, J_3) = (1, 1, j)$, with $j$ an odd number and $j \geq 3$ 

Similar to the previous case. Just interchange the 1-2 symmetric vertexes with 
the 1-2 anti-symmetric ones.

\item $(J_1, J_2, J_3) = (j, j, 1)$, with $j \geq 1$ 

\begin{equation}
\begin{array}{rcrl}
\Gamma_+^{(S)} & = & \left( g^{\alpha_1\gamma_1}p^{\beta_1}
+ g^{\beta_1\gamma_1}p^{\alpha_1}
\right) & \left\{ \; c_1 \left(
g^{\alpha_2\beta_2}g^{\alpha_3\beta_3}\cdots g^{\alpha_j\beta_j}
\right)\right. \\
& & & +c_2 \left(
g^{\alpha_2\beta_2}g^{\alpha_3\beta_3}\cdots g^{\alpha_{j-1}\beta_{j-1}}
	p^{\alpha_j}p^{\beta_j}
\right) \\
& & & +\cdots \\
& & & \left. + c_j \left(
p^{\alpha_2}p^{\alpha_3}\cdots p^{\alpha_j}
	p^{\beta_2}p^{\beta_3}\cdots p^{\beta_j}
\right)\right\},
\end{array}
\end{equation}

\begin{equation}
\begin{array}{rcrl} 
\Gamma_+^{(A)} & = & \left( g^{\alpha_1\gamma_1}p^{\beta_1}
- g^{\beta_1\gamma_1}p^{\alpha_1}
\right) & \left\{ \; c_{j+1} \left(
g^{\alpha_2\beta_2}g^{\alpha_3\beta_3}\cdots g^{\alpha_j\beta_j}
\right)\right. \\
& & & +c_{j+2} \left(
g^{\alpha_2\beta_2}g^{\alpha_3\beta_3}\cdots g^{\alpha_{j-1}\beta_{j-1}}
	p^{\alpha_j}p^{\beta_j}
\right) \\
& & & +\cdots \\
& & & \left. + c_{2j} \left(
p^{\alpha_2}p^{\alpha_3}\cdots p^{\alpha_j}
	p^{\beta_2}p^{\beta_3}\cdots p^{\beta_j}
\right)\right\}\\
& & +k^{\gamma_1} & \left\{ c_{2j+1} \left(
g^{\alpha_1\beta_1}g^{\alpha_2\beta_2}\cdots g^{\alpha_j\beta_j}
\right)\right. \\
& & & +c_{2j+2}\left(
g^{\alpha_1\beta_1}g^{\alpha_2\beta_2}\cdots g^{\alpha_{j-1}\beta_{j-1}}
	p^{\alpha_j}p^{\beta_j}
\right) \\
& & & +\cdots \\
& & & \left. +c_{3j+1}\left(
p^{\alpha_1}p^{\alpha_2}\cdots p^{\alpha_j}
	p^{\beta_1}p^{\beta_2}\cdots p^{\beta_j}
\right)\right\},
\end{array}
\end{equation}

\begin{equation}
\begin{array}{rcrl}
\Gamma_-^{(S)} & = & A_+^{\alpha_1\beta_1\gamma_1} & \left\{
\; c_{3j+2} \left(
g^{\alpha_2\beta_2}g^{\alpha_3\beta_3}\cdots g^{\alpha_j\beta_j}
\right)\right. \\
& & & +c_{3j+3} \left(
g^{\alpha_2\beta_2}g^{\alpha_3\beta_3}\cdots g^{\alpha_{j-1}\beta_{j-1}}
	p^{\alpha_j}p^{\beta_j}
\right)\\
& & & +\cdots \\
& & & \left. +c_{4j+1} \left(
p^{\alpha_2}p^{\alpha_3}\cdots p^{\alpha_j}
	p^{\beta_2}p^{\beta_3}\cdots p^{\beta_j}
\right)\right\},
\end{array}
\end{equation}

\begin{equation}
\begin{array}{rcrl}
\Gamma_-^{(A)} & = & A_-^{\alpha_1\beta_1\gamma_1} & \left\{
\; c_{4j+2} \left(
g^{\alpha_2\beta_2}g^{\alpha_3\beta_3}\cdots g^{\alpha_j\beta_j}
\right)\right. \\
& & & +c_{4j+3} \left(
g^{\alpha_2\beta_2}g^{\alpha_3\beta_3}\cdots g^{\alpha_{j-1}\beta_{j-1}}
	p^{\alpha_j}p^{\beta_j}
\right)\\
& & & +\cdots \\
& & & \left. +c_{5j+1} \left(
p^{\alpha_2}p^{\alpha_3}\cdots p^{\alpha_j}
	p^{\beta_2}p^{\beta_3}\cdots p^{\beta_j}
\right)\right\}\\
& & +Q^{\alpha_1\beta_1}k^{\gamma_1} & \left\{
\; c_{5j+2} \left(
g^{\alpha_2\beta_2}g^{\alpha_3\beta_3}\cdots g^{\alpha_j\beta_j}
\right)\right.\\
& & & +c_{5j+3} \left(
g^{\alpha_2\beta_2}g^{\alpha_3\beta_3}\cdots g^{\alpha_{j-1}\beta_{j-1}}
	p^{\alpha_j}p^{\beta_j}
\right)\\
& & & +\cdots \\
& & & \left. +c_{6j+1} \left(
p^{\alpha_2}p^{\alpha_3}\cdots p^{\alpha_j}
	p^{\beta_2}p^{\beta_3}\cdots p^{\beta_j}
\right)\right\}.
\end{array}
\end{equation}

\item $(J_1, J_2, J_3) = (2, 2, 2)$

\begin{equation}
\begin{array}{rcl}
\Gamma_+^{(S)} & = &
\;\;\; c_1 g^{\alpha_1\beta_1}g^{\alpha_2\beta_2}k^{\gamma_1}k^{\gamma_2} \\
& & + c_2 g^{\alpha_1\beta_1}g^{\alpha_2\gamma_1}g^{\beta_2\gamma_2} \\
& & + c_3 g^{\alpha_1\beta_1} \left(
  g^{\alpha_2\gamma_1}p^{\beta_2} 
- g^{\beta_2\gamma_1}p^{\alpha_2}
\right) k^{\gamma_2} \\
& & + c_4 g^{\alpha_1\beta_1}p^{\alpha_2}p^{\beta_2}k^{\gamma_1}k^{\gamma_2} \\
& & + c_5 \left(
  g^{\alpha_1\gamma_1}g^{\alpha_2\gamma_2}p^{\beta_1}p^{\beta_2}
+ g^{\beta_1\gamma_1}g^{\beta_2\gamma_2}p^{\alpha_1}p^{\alpha_2}
\right)\\
& & + c_6 g^{\alpha_1\gamma_1}g^{\beta_1\gamma_2}p^{\alpha_2}p^{\beta_2} \\
& & + c_7 p^{\alpha_1}p^{\alpha_2}p^{\beta_1}p^{\beta_2}
	k^{\gamma_1}k^{\gamma_2},
\end{array}
\end{equation}

\begin{equation}
\begin{array}{rcl}
\Gamma_+^{(A)} & = &
\;\;\; c_8
g^{\alpha_1\beta_1} \left(
  g^{\alpha_2\gamma_1}p^{\beta_2} 
+ g^{\beta_2\gamma_1}p^{\alpha_2}
\right) k^{\gamma_2} \\
& & +c_9 \left(
  g^{\alpha_1\gamma_1}g^{\alpha_2\gamma_2}p^{\beta_1}p^{\beta_2}
- g^{\beta_1\gamma_1}g^{\beta_2\gamma_2}p^{\alpha_1}p^{\alpha_2}
\right) \\
& & + c_{10} \left(
  g^{\alpha_1\gamma_1}p^{\beta_1}
+ g^{\beta_1\gamma_1}p^{\alpha_1}
\right) p^{\alpha_2}p^{\beta_2}k^{\gamma_2},
\end{array}
\end{equation}

\begin{equation}
\begin{array}{rcl}
\Gamma_-^{(S)} & = &
\;\;\; c_{11} g^{\alpha_1\beta_1}A_-^{\alpha_2\beta_2\gamma_1}k^{\gamma_2}\\
& & +c_{12} g^{\alpha_1\beta_1}Q^{\alpha_2\beta_2}k^{\gamma_1}k^{\gamma_2}\\
& & +c_{13}A_-^{\alpha_1\beta_1\gamma_1}\left(
p^{\alpha_2}g^{\beta_2\gamma_2}
	-p^{\beta_2}g^{\alpha_2\gamma_2}
\right) \\
& & +c_{14} A_+^{\alpha_1\beta_1\gamma_1}\left(
p^{\alpha_2}g^{\beta_2\gamma_2}
	+p^{\beta_2}g^{\alpha_2\gamma_2}
\right)\\
& & +c_{15} Q^{\alpha_1\beta_1}
	p^{\alpha_2}p^{\beta_2}k^{\gamma_1}k^{\gamma_2},
\end{array}
\end{equation}

\begin{equation}
\begin{array}{rcl}
\Gamma_-^{(A)} & = &
\;\;\; c_{16} g^{\alpha_1\beta_1}\left(
  Q^{\alpha_2\gamma_1}g^{\beta_2\gamma_2}
+ Q^{\beta_2\gamma_1}g^{\alpha_2\gamma_2}
\right)\\ 
& & +c_{17} g^{\alpha_1\beta_1}A_+^{\alpha_2\beta_2\gamma_1}k^{\gamma_2}\\
& & +c_{18} A_-^{\alpha_1\beta_1\gamma_1}\left(
p^{\alpha_2}g^{\beta_2\gamma_2}
	+p^{\beta_2}g^{\alpha_2\gamma_2}
\right)\\
& & +c_{19} A_+^{\alpha_1\beta_1\gamma_1}p^{\alpha_2}p^{\beta_2}k^{\gamma_2}.
\end{array}
\end{equation}

\item $(J_1, J_2, J_3) = (2, 2, 3)$

\begin{equation}
\begin{array}{rcl}
\Gamma_+^{(S)} & = & 
\;\;\; c_1 g^{\alpha_1\beta_1}\left(
 g^{\alpha_2\gamma_1}p^{\beta_2}
+g^{\beta_2\gamma_1}p^{\alpha_2}
\right)k^{\gamma_2}k^{\gamma_3}\\
& & +c_2 g^{\alpha_1\gamma_1}g^{\beta_1\gamma_2}\left(
 g^{\alpha_2\gamma_3}p^{\beta_2}
+g^{\beta_2\gamma_2}p^{\alpha_2}
\right)\\
& & +c_3 \left(
 g^{\alpha_1\gamma_1}g^{\alpha_2\gamma_2}p^{\beta_1}p^{\beta_2}
-g^{\beta_1\gamma_1}g^{\beta_2\gamma_2}p^{\alpha_1}p^{\alpha_2}
\right) k^{\gamma_3}\\
& & +c_4 \left(
 g^{\alpha_1\gamma_1}p^{\beta_1}
+g^{\beta_1\gamma_1}p^{\alpha_1}
\right)p^{\alpha_2}p^{\beta_2}k^{\gamma_2}k^{\gamma_3},
\end{array}
\end{equation}

\begin{equation}
\begin{array}{rcl}
\Gamma_+^{(A)} & = &
\;\;\; c_5 g^{\alpha_1\beta_1}g^{\alpha_2\beta_2}
    k^{\gamma_1}k^{\gamma_2}k^{\gamma_3} \\
& & +c_6 g^{\alpha_1\beta_1}g^{\alpha_2\gamma_1}g^{\beta_2\gamma_2}
    k^{\gamma_3} \\
& & +c_7 g^{\alpha_1\beta_1}\left(
 g^{\alpha_2\gamma_1}p^{\beta_2}
-g^{\beta_2\gamma_1}p^{\alpha_2}
\right)k^{\gamma_2}k^{\gamma_3}\\
& & +c_8 g^{\alpha_1\beta_1}p^{\alpha_2}p^{\beta_2}
    k^{\gamma_1} k^{\gamma_2} k^{\gamma_3}\\
& & +c_9 g^{\alpha_1\gamma_1}g^{\beta_1\gamma_2}\left(
 g^{\alpha_2\gamma_3}p^{\beta_2}
-g^{\beta_2\gamma_2}p^{\alpha_2}
\right)\\
& & +c_{10} \left(
 g^{\alpha_1\gamma_1}g^{\alpha_2\gamma_2}p^{\beta_1}p^{\beta_2}
+g^{\beta_1\gamma_1}g^{\beta_2\gamma_2}p^{\alpha_1}p^{\alpha_2}
\right) k^{\gamma_3}\\
& & +c_{11} g^{\alpha_1\gamma_1}g^{\beta_1\gamma_2}
  p^{\alpha_2}p^{\beta_2}k^{\gamma_3}\\
& & +c_{12} p^{\alpha_1}p^{\alpha_2}p^{\beta_1}p^{\beta_2}
	k^{\gamma_1}k^{\gamma_2}k^{\gamma_3},
\end{array}
\end{equation}

\begin{equation}
\begin{array}{rcl}
\Gamma_-^{(S)} & = &
\;\;\; c_{13} g^{\alpha_1\beta_1}A_+^{\alpha_1\beta_2\gamma_1}
	k^{\gamma_2}k^{\gamma_3}\\
& & +c_{14} g^{\alpha_1\beta_1}\left(
 Q^{\alpha_2\gamma_1}g^{\beta_2\gamma_2}
+Q^{\beta_2\gamma_1}g^{\alpha_2\gamma_2}
\right)k^{\gamma_3} \\
& & +c_{15} A_+^{\alpha_1\beta_1\gamma_1}
	g^{\alpha_2\gamma_2}g^{\beta_2\gamma_3}\\
& & +c_{16} A_-^{\alpha_1\beta_1\gamma_1}\left(
 g^{\alpha_2\gamma_2}p^{\beta_2}
+g^{\beta_2\gamma_2}p^{\alpha_2}
\right)k^{\gamma_3} \\
& & +c_{17} A_+^{\alpha_1\beta_1\gamma_1}
	p^{\alpha_2}p^{\beta_2}k^{\gamma_2}k^{\gamma_3},
\end{array}
\end{equation}

\begin{equation}
\begin{array}{rcl}
\Gamma_-^{(A)} & = &
\;\;\; c_{18} g^{\alpha_1\beta_1}A_-^{\alpha_2\beta_2\gamma_1}
	k^{\gamma_2}k^{\gamma_3}\\
& & +c_{19} g^{\alpha_1\beta_1}Q^{\alpha_2\beta_2}
	k^{\gamma_1}k^{\gamma_2}k^{\gamma_3} \\
& & +c_{20} A_-^{\alpha_1\beta_1\gamma_1}
	g^{\alpha_2\gamma_2}g^{\beta_2\gamma_3}\\
& & +c_{21} A_-^{\alpha_1\beta_1\gamma_1}\left(
 g^{\alpha_2\gamma_2}p^{\beta_2}
-g^{\beta_2\gamma_2}p^{\alpha_2}
\right)k^{\gamma_3}\\
& & +c_{22} A_+^{\alpha_1\beta_1\gamma_1}\left(
 g^{\alpha_2\gamma_2}p^{\beta_2}
+g^{\beta_2\gamma_2}p^{\alpha_2}
\right)k^{\gamma_3}\\
& & +c_{23} Q^{\alpha_1\beta_1}p^{\alpha_2}p^{\beta_2}
	k^{\gamma_1}k^{\gamma_2}k^{\gamma_3}.
\end{array}
\end{equation}

\item $(J_1, J_2, J_3) = (2, 2, j)$, with $j$ an even number and $j \geq 4$

\begin{equation}
\begin{array}{rcl}
\Gamma_+^{(S)} & = &
\;\;\; c_1 g^{\alpha_1\beta_1}g^{\alpha_2\beta_2}\left(
k^{\gamma_1}k^{\gamma_2}\cdots k^{\gamma_j}
\right)\\
& & +c_2 g^{\alpha_1\beta_1}
g^{\alpha_2\gamma_1}g^{\beta_2\gamma_2}\left(
k^{\gamma_3}k^{\gamma_4}\cdots k^{\gamma_j}
\right)\\
& & +c_3 g^{\alpha_1\beta_1}\left(
 g^{\alpha_2\gamma_1}p^{\beta_2}
-g^{\beta_2\gamma_1}p^{\alpha_2}
\right)\left(
k^{\gamma_2}k^{\gamma_3}\cdots k^{\gamma_j}
\right)\\
& & +c_4 g^{\alpha_1\beta_1}p^{\alpha_2}p^{\beta_2}\left(
	k^{\gamma_1}k^{\gamma_2}\cdots k^{\gamma_j}
\right)\\
& & +c_5 
g^{\alpha_1\gamma_1}g^{\alpha_2\gamma_2}g^{\beta_1\gamma_3}g^{\beta_2\gamma_4}
\left(
k^{\gamma_5}k^{\gamma_6}\cdots k^{\gamma_j}
\right)\\
& & +c_6 g^{\alpha_1\gamma_1}g^{\beta_1\gamma_2}\left(
 g^{\alpha_2\gamma_3}p^{\beta_2}
-g^{\beta_2\gamma_3}p^{\alpha_2}
\right)\left(
k^{\gamma_4}k^{\gamma_5}\cdots k^{\gamma_j}
\right)\\
& & +c_7 g^{\alpha_1\gamma_1}g^{\beta_1\gamma_2}
	p^{\alpha_2}p^{\beta_2}\left(
k^{\gamma_3}k^{\gamma_4}\cdots k^{\gamma_j}
\right)\\
& & +c_8 \left(
 g^{\alpha_1\gamma_1}g^{\alpha_2\gamma_2}p^{\beta_1}p^{\beta_2}
+g^{\beta_1\gamma_1}g^{\beta_2\gamma_2}p^{\alpha_1}p^{\alpha_2}
\right)\left(
k^{\gamma_3}k^{\gamma_4}\cdots k^{\gamma_j}
\right)\\
& & +c_9 p^{\alpha_1}p^{\alpha_2}p^{\beta_1}p^{\beta_2}\left(
k^{\gamma_1}k^{\gamma_2}\cdots k^{\gamma_j}
\right),
\end{array}
\end{equation}

\begin{equation}
\begin{array}{rcl}
\Gamma_+^{(A)} & = &
\;\;\; c_{10} g^{\alpha_1\beta_1}\left(
 g^{\alpha_2\gamma_1}p^{\beta_2}
+g^{\beta_2\gamma_1}p^{\alpha_2}
\right)\left(
k^{\gamma_2}k^{\gamma_3}\cdots k^{\gamma_j}
\right)\\
& & +c_{11} g^{\alpha_1\gamma_1}g^{\beta_1\gamma_2}\left(
 g^{\alpha_2\gamma_3}p^{\beta_2}
+g^{\beta_2\gamma_3}p^{\alpha_2}
\right)\left(
k^{\gamma_4}k^{\gamma_5}\cdots k^{\gamma_j}
\right)\\
& & +c_{12} \left(
 g^{\alpha_1\gamma_1}g^{\alpha_2\gamma_2}p^{\beta_1}p^{\beta_2}
-g^{\beta_1\gamma_1}g^{\beta_2\gamma_2}p^{\alpha_1}p^{\alpha_2}
\right)\left(
k^{\gamma_3}k^{\gamma_4}\cdots k^{\gamma_j}
\right)\\
& & +c_{13} \left(
 g^{\alpha_1\gamma_1}p^{\beta_1}
+g^{\beta_1\gamma_1}p^{\alpha_1}
\right)p^{\alpha_2}p^{\beta_2}\left(
k^{\gamma_2}k^{\gamma_3}\cdots k^{\gamma_j}
\right),
\end{array}
\end{equation}

\begin{equation}
\begin{array}{rcl}
\Gamma_-^{(S)} & = &
\;\;\; c_{14} g^{\alpha_1\beta_1}A_-^{\alpha_2\beta_2\gamma_1}\left(
	k^{\gamma_2}k^{\gamma_3}\cdots k^{\gamma_j}
\right)\\
& & +c_{15} g^{\alpha_1\beta_1}Q^{\alpha_2\beta_2}\left(
k^{\gamma_1}k^{\gamma_2}\cdots k^{\gamma_j}
\right)\\
& & +c_{16} A_-^{\alpha_1\beta_1\gamma_1}
	g^{\alpha_2\gamma_2}g^{\beta_2\gamma_3}\left(
	k^{\gamma_4}k^{\gamma_5}\cdots k^{\gamma_j}
\right)\\
& & +c_{17} A_+^{\alpha_1\beta_1\gamma_1}\left(
 g^{\alpha_2\gamma_2}p^{\beta_2}
+g^{\beta_2\gamma_2}p^{\alpha_2}
\right)\left(
k^{\gamma_3}k^{\gamma_4}\cdots k^{\gamma_j}
\right)\\
& & +c_{18} A_-^{\alpha_1\beta_1\gamma_1}\left(
 g^{\alpha_2\gamma_2}p^{\beta_2}
-g^{\beta_2\gamma_2}p^{\alpha_2}
\right)\left(
k^{\gamma_3}k^{\gamma_4}\cdots k^{\gamma_j}
\right)\\
& & +c_{19} Q^{\alpha_1\beta_1}p^{\alpha_2}p^{\beta_2}\left(
k^{\gamma_1}k^{\gamma_2}\cdots k^{\gamma_j}
\right),
\end{array}
\end{equation}

\begin{equation}
\begin{array}{rcl}
\Gamma_-^{(A)} & = &
\;\;\; c_{20} g^{\alpha_1\beta_1}A_+^{\alpha_2\beta_2\gamma_1}\left(
	k^{\gamma_2}k^{\gamma_3}\cdots k^{\gamma_j}
\right)\\
& & +c_{21} g^{\alpha_1\beta_1}\left(
 Q^{\alpha_2\gamma_1}g^{\beta_2\gamma_2}
+Q^{\beta_2\gamma_1}g^{\alpha_2\gamma_2}
\right)\left(
k^{\gamma_3}k^{\gamma_4}\cdots k^{\gamma_j}
\right)\\
& & +c_{22} A_+^{\alpha_1\beta_1\gamma_1}
	g^{\alpha_1\gamma_2}g^{\beta_2\gamma_3}\left(
	k^{\gamma_4}k^{\gamma_5}\cdots k^{\gamma_j}
\right)\\
& & +c_{23} A_-^{\alpha_1\beta_1\gamma_1}\left(
 g^{\alpha_2\gamma_2}p^{\beta_2}
+g^{\beta_2\gamma_2}p^{\alpha_2}
\right)\left(
k^{\gamma_3}k^{\gamma_4}\cdots k^{\gamma_j}
\right)\\
& & +c_{24} A_+^{\alpha_1\beta_1\gamma_1}
	p^{\alpha_2}p^{\beta_2}\left(
	k^{\gamma_2}k^{\gamma_3}\cdots k^{\gamma_j}
\right)\\
& & +c_{25} \left(
 Q^{\alpha_1\gamma_1}g^{\beta_1\gamma_2}
+Q^{\beta_1\gamma_1}g^{\alpha_1\gamma_2}
\right)g^{\alpha_2\gamma_3}g^{\beta_2\gamma_4}\left(
k^{\gamma_5}k^{\gamma_6}\cdots k^{\gamma_j}
\right).
\end{array}
\end{equation}

\item $(J_1, J_2, J_3) = (2, 2, j)$, with $j$ an odd number and $j \geq 5$

Similar to the previous case, with 
$\Gamma_\pm^{(S)}\longleftrightarrow \Gamma_\pm^{(A)}$.

\item $(J_1, J_2, J_3) = (j, j, 2)$, with $j \geq 3$

\begin{equation}
\begin{array}{rcl}
\Gamma_+^{(S)} & = & 
\left(g^{\alpha_1\gamma_1}g^{\alpha_2\gamma_2}p^{\beta_1}p^{\beta_2}
     +g^{\beta_1\gamma_1}g^{\beta_2\gamma_2}p^{\alpha_1}p^{\alpha_2}
\right) \times \\
& & \left\{c_1 \left(
g^{\alpha_3\beta_3}g^{\alpha_4\beta_4}\cdots g^{\alpha_j\beta_j}
\right)\right.\\
& &  +c_2 \left(
g^{\alpha_3\beta_3}g^{\alpha_4\beta_4}\cdots g^{\alpha_{j-1}\beta_{j-1}}
	p^{\alpha_j}p^{\beta_j}
\right)\\
& &  +\cdots \\
& &  \left. +c_{j-1}\left(
p^{\alpha_3}p^{\alpha_4}\cdots p^{\alpha_j}
	p^{\beta_3}p^{\beta_4}\cdots p^{\beta_j}
\right)\right\}\\
& & +g^{\alpha_1\gamma_1}g^{\beta_1\gamma_2}\times\\
& & \left\{c_{j}\left(
g^{\alpha_2\beta_2}g^{\alpha_3\beta_3}\cdots g^{\alpha_j\beta_j}
\right)\right.\\
& & +c_{j+1}\left(
g^{\alpha_2\beta_2}g^{\alpha_3\beta_3}\cdots g^{\alpha_{j-1}\beta_{j-1}}
	p^{\alpha_j}p^{\beta_j}
\right)\\
& & +\cdots \\
& & \left. +c_{2j-1}\left(
p^{\alpha_2}p^{\alpha_3}\cdots p^{\alpha_j}
	p^{\beta_2}p^{\beta_3}\cdots p^{\beta_j}
\right)\right\}\\
& & \left(g^{\alpha_1\gamma_1}p^{\beta_1}
         +g^{\beta_1\gamma_1}p^{\alpha_1}\right)k^{\gamma_2}\times\\
& & \left\{c_{2j}\left(
g^{\alpha_2\beta_2}g^{\alpha_3\beta_3}\cdots g^{\alpha_j\beta_j}
\right)\right.\\
& & +c_{2j+1}\left(
g^{\alpha_2\beta_2}g^{\alpha_3\beta_3}\cdots g^{\alpha_{j-1}\beta_{j-1}}
	p^{\alpha_j}p^{\beta_j}
\right)\\
& & \cdots \\
& & \left. +c_{3j-1}\left(
p^{\alpha_2}p^{\alpha_3}\cdots p^{\alpha_j}
	p^{\beta_2}p^{\beta_3}\cdots p^{\beta_j}
\right)\right\}\\
& & + c_{3j}k^{\gamma_1}k^{\gamma_2}\left(
g^{\alpha_1\beta_1}g^{\alpha_2\beta_2}\cdots g^{\alpha_j\beta_j}
\right)\\
& & + c_{3j+1}k^{\gamma_1}k^{\gamma_2}\left(
p^{\alpha_1}p^{\alpha_2}\cdots p^{\alpha_j}
	p^{\beta_1}p^{\beta_2}\cdots p^{\beta_j}
\right),
\end{array}
\end{equation}

\begin{equation}
\begin{array}{rcl}
\Gamma_+^{(A)} & = &
\left(g^{\alpha_1\gamma_1}g^{\alpha_2\gamma_2}p^{\beta_1}p^{\beta_2}
     -g^{\beta_1\gamma_1}g^{\beta_2\gamma_2}p^{\alpha_1}p^{\alpha_2}
\right) \times \\
& & \left\{c_{3j+2} \left(
g^{\alpha_3\beta_3}g^{\alpha_4\beta_4}\cdots g^{\alpha_j\beta_j}
\right)\right.\\
& & +c_{3j+3} \left(
g^{\alpha_3\beta_3}g^{\alpha_4\beta_4}\cdots g^{\alpha_{j-1}\beta_{j-1}}
	p^{\alpha_j}p^{\beta_j}
\right)\\
& & +\cdots \\
& & \left. +c_{4j}\left(
p^{\alpha_3}p^{\alpha_4}\cdots p^{\alpha_j}
	p^{\beta_3}p^{\beta_4}\cdots p^{\beta_j}
\right)\right\}\\
& & \left(g^{\alpha_1\gamma_1}p^{\beta_1}
         -g^{\beta_1\gamma_1}p^{\alpha_1}\right)k^{\gamma_2}\times\\
& & \left\{c_{4j+1}\left(
g^{\alpha_2\beta_2}g^{\alpha_3\beta_3}\cdots g^{\alpha_j\beta_j}
\right)\right.\\
& & +c_{4j+2}\left(
g^{\alpha_2\beta_2}g^{\alpha_3\beta_3}\cdots g^{\alpha_{j-1}\beta_{j-1}}
	p^{\alpha_j}p^{\beta_j}
\right)\\
& & \cdots \\
& & \left. +c_{5j}\left(
p^{\alpha_2}p^{\alpha_3}\cdots p^{\alpha_j}
	p^{\beta_2}p^{\beta_3}\cdots p^{\beta_j}
\right)\right\},
\end{array}
\end{equation}

\begin{equation}
\begin{array}{rcl}
\Gamma_-^{(S)} & = &
A_-^{\alpha_1\beta_1\gamma_1}\left(
 p^{\alpha_2}g^{\beta_2\gamma_2}
-p^{\beta_2}g^{\alpha_2\gamma_2}
\right)\times\\
& & \left\{c_{5j+1}\left(
g^{\alpha_3\beta_3}g^{\alpha_4\beta_4}\cdots g^{\alpha_j\beta_j}
\right)\right.\\
& & +c_{5j+2}\left(
g^{\alpha_3\beta_3}g^{\alpha_4\beta_4}\cdots g^{\alpha_{j-1}\beta_{j-1}}
	p^{\alpha_j}p^{\beta_j}
\right)\\
& & +\cdots \\
& & \left. +c_{6j-1}\left(
p^{\alpha_3}p^{\alpha_4}\cdots p^{\alpha_j}
	p^{\beta_3}p^{\beta_4}\cdots p^{\beta_j}
\right)\right\}\\
& & +A_+^{\alpha_1\beta_1\gamma_1}\left(
 p^{\alpha_2}g^{\beta_2\gamma_2}
+p^{\beta_2}g^{\alpha_2\gamma_2}
\right)\times\\
& & \left\{c_{6j}\left(
g^{\alpha_3\beta_3}g^{\alpha_4\beta_4}\cdots g^{\alpha_j\beta_j}
\right)\right.\\
& & +c_{6j+1}\left(
g^{\alpha_3\beta_3}g^{\alpha_4\beta_4}\cdots g^{\alpha_{j-1}\beta_{j-1}}
	p^{\alpha_j}p^{\beta_j}
\right)\\
& & +\cdots \\
& & \left. +c_{7j-2}\left(
p^{\alpha_3}p^{\alpha_4}\cdots p^{\alpha_j}
	p^{\beta_3}p^{\beta_4}\cdots p^{\beta_j}
\right)\right\}\\
& & +c_{7j-1}A_-^{\alpha_1\beta_1\gamma_1}k^{\gamma_2}\left(
g^{\alpha_2\beta_2}g^{\alpha_3\beta_3}\cdots g^{\alpha_j\beta_j}
\right)\\
& & +Q^{\alpha_1\beta_1}k^{\gamma_1}k^{\gamma_2}\times\\
& & \left\{c_{7j}\left(
g^{\alpha_2\beta_2}g^{\alpha_3\beta_3}\cdots g^{\alpha_j\beta_j}
\right)\right.\\
& & +c_{7j+1}\left(
g^{\alpha_2\beta_2}g^{\alpha_3\beta_3}\cdots g^{\alpha_{j-1}\beta_{j-1}}
	p^{\alpha_j}p^{\beta_j}
\right)\\
& & +\cdots \\
& & \left. +c_{8j-1}\left(
p^{\alpha_2}p^{\alpha_3}\cdots p^{\alpha_j}
	p^{\beta_2}p^{\beta_3}\cdots p^{\beta_j}
\right)\right\},
\end{array}
\end{equation}

\begin{equation}
\begin{array}{rcl}
\Gamma_-^{(A)} & = &
A_-^{\alpha_1\beta_1\gamma_1}\left(
 p^{\alpha_2}g^{\beta_2\gamma_2}
+p^{\beta_2}g^{\alpha_2\gamma_2}
\right)\times\\
& & \left\{c_{8j}\left(
g^{\alpha_3\beta_3}g^{\alpha_4\beta_4}\cdots g^{\alpha_j\beta_j}
\right)\right.\\
& & +c_{8j+1}\left(
g^{\alpha_3\beta_3}g^{\alpha_4\beta_4}\cdots g^{\alpha_{j-1}\beta_{j-1}}
	p^{\alpha_j}p^{\beta_j}
\right)\\
& & +\cdots \\
& & \left. +c_{9j-2}\left(
p^{\alpha_3}p^{\alpha_4}\cdots p^{\alpha_j}
	p^{\beta_3}p^{\beta_4}\cdots p^{\beta_j}
\right)\right\}\\
& & +A_+^{\alpha_1\beta_1\gamma_1}k^{\gamma_2}\times\\
& & \left\{c_{9j-1}\left(
g^{\alpha_2\beta_2}g^{\alpha_3\beta_3}\cdots g^{\alpha_j\beta_j}
\right)\right.\\
& & +c_{9j}\left(
g^{\alpha_2\beta_2}g^{\alpha_3\beta_3}\cdots g^{\alpha_{j-1}\beta_{j-1}}
	p^{\alpha_j}p^{\beta_j}
\right)\\
& & +\cdots \\
& & \left. +c_{10j-2}\left(
p^{\alpha_2}p^{\alpha_3}\cdots p^{\alpha_j}
	p^{\beta_2}p^{\beta_3}\cdots p^{\beta_j}
\right)\right\}\\
& & +c_{10j-1} \left(Q^{\alpha_1\gamma_1}g^{\beta_1\gamma_2}
               +Q^{\beta_1\gamma_1}g^{\alpha_1\gamma_2}\right)
\left(g^{\alpha_2\beta2}g^{\alpha_3\beta_3}\cdots g^{\alpha_j\beta_j}\right)
\end{array}
\end{equation}

\end{itemize}
and so on.

We need to consider three cases:
\begin{enumerate}

\item Both of the two identical particles are internal lines

According to Eq.~{(\ref{eq:bs})}, one can obtain

\begin{equation}
\Gamma_\pm = \Gamma_\pm^{(S)} + (p_1^2-p_2^2)\Gamma_\pm^{(A)}
\end{equation}
with the coefficients satisfying
\begin{equation}
c_i = c_i\left(p_1^2+p_2^2, (p_1^2-p_2^2)^2, p_1\cdot p_2 \right).
\end{equation}

\item One is on shell, another an internal line

The factor $(p_1^2-W^2)$ or $(W^2-p_2^2)$ will eliminate the denominator
of the propagator for the off shell particle. This make the contribution
from  $\Gamma_\pm^{(A)}$ has no pole at the particle's mass,
so that it can be absorbed K. S. freely into background amplitudes.
\begin{equation}
\Gamma_\pm = \Gamma_\pm^{(S)},
\end{equation}
\begin{equation}
c_i = c_i\left(p_1^2, p_1\cdot p_2 \right).
\end{equation}

\item Both of the two identical particles are on shell

Since $p_1^2-p_2^2 = 0$, we have
\begin{equation}
\Gamma_\pm = \Gamma_\pm^{(S)}
\end{equation}
with the coefficients
\begin{equation}
c_i = c_i\left(p_1\cdot p_2 \right).
\end{equation}

\end{enumerate}

From Eqs.~{(\ref{eq:00j.1},\ref{eq:00j.2})} one can infer that a 
particle decaying into two identical spin-$0$ particles must has a
even spin. If parity is conserved, the parity of such a particle must be +1.
For example, $\rho^0,\; \eta,\; \eta\prime,\; \omega,\; \phi,\; a_1,\; f_1$ 
etc. will not decay into two neutral pions\footnote{$\rho^0\rightarrow 2 \pi^0$
also violates (approximate) isospin symmetry. Since boson symmetry is an  
exact symmetry, this process is absolutely forbidden.}.

Examples for incorporating boson symmetry in 4-leg vertexes can be found
in Sec.~\ref{sec:bk} and Ref.~\cite{Zhua23pi}.

\section{Covariant helicity amplitudes for two-body decays}\label{sec:2bd}
Helicity amplitudes for two body-decays can be write down directly using
the wave functions and vertexes given in previous sections. Such amplitudes
can be calculated in arbitrary reference frame. Especially they can be 
calculated in laboratory frame so that no Lorentz transformation is needed. 
However, one might still favor amplitudes in center of mass frame(CM frame). 
We will give some explicit results calculated in the rest frame of parent
particles in this section. 

Suppose a spin-$J$ particle with momentum $p$ decays into a spin-$s$ and
a spin-$\sigma$ particle with momentum $q$ and $k$. The helicity amplitude
of such a process is
\begin{equation}
\begin{array}{rcl}
{\cal M}_{\delta\lambda\nu}(p,q,k) & \equiv &
\langle \roarrow{q},s,\lambda;\roarrow{k},\sigma,\nu|{\bf S}
|\roarrow{p},J,\delta\rangle\\
& = &
\sum\limits_{\lambda^{'}\nu^{'}}
D^{s\ast}_{\lambda^{'}\lambda}\left(
	W(L^{-1}(p),q)
	\right)
D^{\sigma\ast}_{\nu^{'}\nu}\left(
	W(L^{-1}(p),k)
	\right)
\langle \roarrow{q}^{'},s,\lambda^{'};\roarrow{k}^{'},\sigma^{'},\nu|{\bf S}
|\roarrow{0},J,\delta\rangle
\end{array}
\end{equation}
with
\begin{eqnarray}
q^{'\alpha} = {L^{-1 \alpha}}_\beta(p)q^\beta,\\
k^{'\alpha} = {L^{-1 \alpha}}_\beta(p)k^\beta.
\end{eqnarray}
Here Eq.~{(\ref{eq:lrt})} has been used. Notice that we should use
$L(p)$ defined in Eq.~{(\ref{eq:lh})}.

From Eqs.~{(\ref{eq:rot1},\ref{eq:rot2})} one can rotate $\roarrow{q}^{'}$ to 
the direction of $z$-axis:
\begin{equation}\label{eq:rel}
{\cal M}_{\delta\lambda\nu}(p,q,k)  =
\sum\limits_{\lambda^{'}\nu^{'}}
D^{s\ast}_{\lambda^{'}\lambda}\left(
	W(L^{-1}(p),q)
	\right)
D^{\sigma\ast}_{\nu^{'}\nu}\left(
	W(L^{-1}(p),k)
	\right)
D^{J\ast}_{\lambda^{'}-\nu{'},\delta}(\varphi,\vartheta,0) 
F_{\lambda^{'}\nu^{'}},
\end{equation}
where $(\vartheta, \varphi)$ is the direction of $\roarrow{q}^{'}$.
$F_{\lambda\nu}$ is the helicity amplitude
in the rest frame of the parent particle,
\begin{equation}
F_{\lambda\nu} = \langle q,s,\lambda; k,\sigma,\nu|{\bf S}|
	p,J,\lambda-\nu \rangle _{CM} .
\end{equation}
$p,q,k$ have been redefined to their CM frame values in the above equation.  
We follow the convention of Chung~\cite{ChungHA95} in this section:
\begin{equation}
\begin{array}{cclccr}
(p^\alpha) & = & ( W; & 0, & 0, & 0 ),\\
(q^\alpha) & = & ( q_0; & 0, & 0, & r/2 ),\\
(k^\alpha) & = & ( k_0; & 0, & 0, & -r/2 ).
\end{array}
\end{equation}

Eq.~{(\ref{eq:rel})} is the relation between helicity amplitudes in
laboratory frame and those in CM frame. It can be derived in a
alternative way by writing down the explicit expressions for these 
amplitudes and use Lorentz transformation properties of wave functions in 
Eq.~{(\ref{eq:wltr})}.

The masses of the daughter particles are $m$ and $\mu$,
\begin{eqnarray}
W & = & q_0 + k_0 , \\
q_0 & = & \sqrt{m^2 + \frac{r^2}{4}},\\
k_0 & = & \sqrt{\mu^2 + \frac{r^2}{4}}.
\end{eqnarray}
The corresponding space reflection parity of the three particles are
$\eta_J$, $\eta_s$ and $\eta_\sigma$.  

Parity conserving helicity amplitudes in CM frame 
satisfy~\cite{ChungTA93,ChungSF71}
\begin{equation}
F_{\lambda\nu} = \eta_J \eta_s \eta_\sigma (-1)^{J-s-\sigma} 
F_{-\lambda,-\nu}.
\end{equation}

If the two daughter particles are identical, 
one has~\cite{JacobH59,ChungTA93,ChungSF71}
\begin{equation}
F_{\lambda\nu} = (-1)^J F_{\nu\lambda}.
\end{equation}

Some explicit results for $F_{\lambda\nu}$ are listed below.
We assume space reflection symmetry in all processes.

\begin{itemize}

\item Spin-1 $\longrightarrow$ spin-0 + spin-1, 
$\eta_J \eta_s \eta_\sigma = -1$

We should choose pseudo-tensor effective vertexes.
\begin{equation}
F_{\lambda\nu}  =  - F_{-\lambda,-\nu} ;
\end{equation}
\begin{equation}
F_{01}  =  \frac{i}{2} c W r.
\end{equation}
Here $c$ is a scalar.

\item Spin-1 $\longrightarrow$ spin-0 + spin-1, 
$\eta_J \eta_s \eta_\sigma = +1$

The effective vertex should be a tensor.
\begin{equation}
F_{\lambda\nu}  =  + F_{-\lambda,-\nu};
\end{equation}
\begin{eqnarray}
F_{01} & = & -\frac{k_0}{\mu}c_2 + \frac{W}{4\mu}c_1 r^2,\\ 
F_{00} & = & -c_2.
\end{eqnarray}
It can be applied to the process $a_1(1260)\longrightarrow \pi\rho$.

\item Spin-1 $\longrightarrow$ spin-2 + spin-1, 
$\eta_J \eta_s \eta_\sigma = +1$

The effective vertex is a tensor.
\begin{equation}
F_{\lambda\nu}  =  + F_{-\lambda,-\nu};
\end{equation}
\begin{eqnarray}
F_{00} & = & \sqrt{\frac{2}{3}}\frac{k_0 q_0^2}{\mu m^2}c_5 
	+\frac{1}{2\sqrt{6}\mu m^2}\left(c_5 q_0 + c_3 k_0 q_0 W
		-c_2 k_0 W^2 + c_4 q_0 W^2
		\right)r^2
 	+\frac{W}{8\sqrt{6}\mu m^2}\left(c_3 + c_1 W^2 \right)r^4,\\ 
F_{10} & = & \frac{k_0 q_0}{\sqrt{2}\mu m}c_5
	+ \frac{1}{4\sqrt{2}\mu m}\left(c_5 + c_4 W^2
		\right)r^2,\\
F_{01} & = & -\frac{1}{\sqrt{6}}c_5
	-\frac{W^2}{2\sqrt{6}m^2}c_2 r^2,\\ 
F_{11} & = & - \frac{q_0}{\sqrt{2} m} c_5
	-\frac{W}{2\sqrt{2} m} c_3 r^2,\\
F_{21} & = & -c_5.
\end{eqnarray}
These amplitudes are applicable to the case 
$J/\psi\longrightarrow a_2(1320) \rho$.

\item Spin-1 $\longrightarrow$ spin-4 + spin-1, 
$\eta_J \eta_s \eta_\sigma = +1$

Use the tensor vertexes given in Sec.~{\ref{sec:3v}}.
\begin{equation}
F_{\lambda\nu} = + F_{-\lambda,-\nu};
\end{equation}
\begin{eqnarray}
F_{00} & = & \frac{k_0 q_0 W^2}{\sqrt{70} m^4}c_5 r^2
	+\frac{W^2}{\sqrt{1120} m^4}\left(
		c_5 q_0 + c_3 k_0 q_0 W - c_2 k_0 W^2 + c_4 q_0 W^2
		\right) r^4 
 	+ \frac{W^3}{\sqrt{17920} m^4} \left(c_3 + c_1 W^2
		\right) r^6,\\
F_{01} & = & -\frac{W^2}{\sqrt{280}m^2}c_5 r^2
	-\frac{W^4}{\sqrt{1120}m^4}c_4 r^4,\\
F_{10} & = & \frac{k_0 q_0 W^2}{4\sqrt{7}m^3}c_5 r^2
	+\frac{W^2}{16\sqrt{7}m^3}\left(c_5 +c_4 W^2
		\right) r^4,\\
F_{11} & = & -\frac{q_0 W^2}{4\sqrt{7}m^3}c_5 r^2
	-\frac{W^3}{16\sqrt{7}m^3}c_3 r^4,\\
F_{21} & = & -\frac{W^2}{4\sqrt{7}m^2}c_5 r^2.
\end{eqnarray}

\item Spin-0 $\longrightarrow$ spin-1 + spin-1, 
$\eta_J \eta_s \eta_\sigma = -1$

The effective vertex is a pseudo scalar. 
\begin{equation}
F_{\lambda\nu} = - F_{-\lambda,-\nu};
\end{equation}
\begin{equation}
F_{11} = i g W r.
\end{equation}
These amplitudes automatically satisfy
\begin{equation}
F_{\lambda\nu}  =  F_{\nu\lambda}.
\end{equation}

\item Spin-0 $\longrightarrow$ spin-1 + spin-1, 
$\eta_J \eta_s \eta_\sigma = +1$

The vertex is a tensor.
\begin{equation}
F_{\lambda\nu} = + F_{-\lambda,-\nu};
\end{equation}
\begin{eqnarray}
F_{00} & = & -\frac{k_0 q_0}{m \mu}c_2
	-\frac{1}{4m\mu}\left(c_2 + c_1 W^2
		\right)r^2,\\
F_{11} & = & c_2.
\end{eqnarray}

\item Spin-0 $\longrightarrow$ spin-0 + spin-1, 
$\eta_J \eta_s \eta_\sigma = -1$

The vertex should be a tensor.
\begin{equation}
F_{00} = -\frac{W}{2\mu}c .
\end{equation}

\item Spin-2 $\longrightarrow$ spin-0 + spin-0, 
$\eta_J \eta_s \eta_\sigma = +1$

The vertex should be a tensor,
\begin{equation}
F_{00} = -\frac{1}{\sqrt{24}}c r^2 .
\end{equation}
It can be applied to the decay $f_2(1270)\longrightarrow \pi^+ \pi^-$.

\item Spin-2 $\longrightarrow$ two identical spin-1 particles, 
$\eta_J = +1$

The vertex should satisfy boson symmetry.
\begin{eqnarray}
F_{\lambda\nu} & = & + F_{-\lambda,-\nu},\\
F_{\lambda\nu} & = & F_{\nu\lambda}; 
\end{eqnarray}
\begin{eqnarray}
F_{00} & = & \sqrt{\frac{2}{3}}\frac{q_0^2}{m^2}c_4
	+\frac{1}{\sqrt{6}m^2}\left(
		\frac{c_2}{2}q_0^2 - c_3 q_0 W
		\right) r^2
	+\frac{1}{\sqrt{384}m^2}\left(
		c_2 + c_1 W^2
		\right) r^4,\\
F_{01} & = & \frac{q_0}{\sqrt{2}m}c_4
	-\frac{W}{4\sqrt{2}m}c_3 r^2,\\
F_{1,-1} & = & c_4 ,\\
F_{11} & = & \frac{1}{\sqrt{6}}c_4
	-\frac{1}{2\sqrt{6}}c_2 r^2 .
\end{eqnarray}

\item Spin-2 $\longrightarrow$ spin-2 + spin-0, 
$\eta_J \eta_s \eta_\sigma = +1$

The effective vertex should be a tensor.
\begin{equation}
F_{\lambda\nu} = + F_{-\lambda,-\nu};
\end{equation}
\begin{eqnarray}
F_{00} & = & \frac{1}{3}\left(1+\frac{2q_0^2}{m^2}\right)c_3
	-\frac{q_0 W}{6 m^2}c_2 r^2
	+\frac{W^2}{24 m^2}c_1 r^4 ,\\
F_{10} & = & \frac{q_0}{m}c_3 - \frac{W}{8 m}c_2 r^2,\\
F_{20} & = & c_3 .
\end{eqnarray}

\item Spin-4 $\longrightarrow$ two identical spin-1 particles, 
$\eta_J = +1$

The vertex should satisfy boson symmetry.
\begin{eqnarray}
F_{\lambda\nu} & = & + F_{-\lambda,-\nu},\\
F_{\lambda\nu} & = & F_{\nu\lambda}; 
\end{eqnarray}
\begin{eqnarray}
F_{00} & = & -\frac{q_0^2}{\sqrt{70}m^2}c_4 r^2
	-\frac{1}{\sqrt{1120}m^2}\left( c_2 q_0^2 + 
		2 c_3 q_0 W
		\right)r^4
	-\frac{1}{\sqrt{17920}m^2}\left(c_2 + c_1 W^2
		\right) r^6 ,\\
F_{01} & = & -\frac{q_0}{4\sqrt{7}m}c_4 r^2
	-\frac{W}{16\sqrt{7}m}c_3 r^4 ,\\
F_{1,-1} & = & -\frac{1}{4\sqrt{7}}c_4 r^2,\\
F_{11} &= & -\frac{1}{\sqrt{280}}c_4 r^2
	+\frac{1}{\sqrt{1120}}c_2 r^4 .
\end{eqnarray}

\item Spin-4 $\longrightarrow$ spin-2 + spin-0, 
$\eta_J \eta_s \eta_\sigma = +1$

The effective vertex is a tensor.
\begin{equation}
F_{\lambda\nu} = + F_{-\lambda,-\nu};
\end{equation}
\begin{eqnarray}
F_{00} & = & \frac{1}{2\sqrt{105}}\left(1+ \frac{2q_0^2}{m^2}
		\right)c_3 r^2
	-\frac{W q_0}{4\sqrt{105}m^2}c_2 r^4
	+\frac{W^2}{16\sqrt{105}m^2}c_1 r^6 ,\\
F_{10} & = & \frac{q_0}{\sqrt{56}m}c_3 r^2
	-\frac{W}{\sqrt{3584}m}c_2 r^4 ,\\
F_{20} & = & \frac{1}{4\sqrt{7}}c_3 r^2 .
\end{eqnarray}

\item Spin-1 $\longrightarrow$ spin-0 + spin-0, 
$\eta_J \eta_s \eta_\sigma = -1$

The effective vertex should be a vector.
\begin{equation}
F_{00} = -\frac{c}{2}r .
\end{equation}

\end{itemize}

\section{Resonances and backgrounds}\label{sec:bk}
\begin{figure}[htbp]
\centerline{\psfig{figure=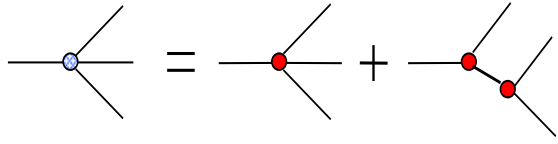,height=0.8in,width=3in}}
\caption{A 4-leg vertex can be divided into an 1PI part 
	and resonance parts}\label{fig:4leg}
\end{figure}

For a process involving more than three particles, we must separate the 
vertexes into one-particle irreducible (1PI) parts and one-particle reducible 
(1PR) parts. Usually 1PI parts are called backgrounds, while 1PR parts are 
called resonances. An example with 4-leg vertex are shown in 
Fig.~{\ref{fig:4leg}}.

\vspace{2.3cm}

\begin{figure}[htbp]
\centerline{\psfig{figure=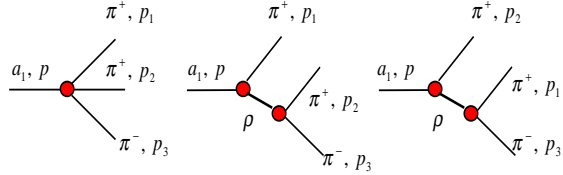,height=1in,width=3in}}
\caption{The 1PI graph and two $\rho$-resonance graphs 
  for the process $a_1\rightarrow \pi^+\pi^+\pi^-$}\label{fig:a13pi}
\end{figure}

Feynman graphs for the process $a_1\rightarrow \pi^+\pi^+\pi^-$ are 
illustrated in Fig.~\ref{fig:a13pi}. We consider only $\rho$ resonance here.
The four-momenta of $a_1$ and the three pions are $p$, $p_1$,
$p_2$ and $p_3$, with
\begin{equation}
p = p_1 + p_2 + p_3.
\end{equation}
The corresponding spin-parities of $a_1, \rho$ and $\pi$ are
$1^-$, $1^-$ and $0^-$\cite{pdg98}.

\vspace{2.3cm}

The background amplitude can be found after
an analysis similar to that in Ref.~\cite{Zhua23pi}\footnote{
The $c_3$ term in Eq.(10) of Ref.~\cite{Zhua23pi} can be dropped 
without introducing K. S.. }:
\begin{eqnarray}
{\cal M}_\lambda^{(b)} & = & e_\mu(p,\lambda) \{
  b_1 (p_1+p_2)^\mu 
+ [(p_1-p_2)\cdot p_3] b_2 (p_1-p_2)^\mu \}, \\
b_i & = & b_i\left( (p_1+p_2)\cdot p_3, [(p_1-p_2)\cdot p_3]^2 \right).
\end{eqnarray}

We use Breit-Wigner factors as approximation to the full propagators 
in Fig.~\ref{fig:a13pi}, i.e., write the propagators of $\rho^0$ as
\begin{equation}
\frac{g^{\alpha\beta}-(p_2+p_3)^\alpha(p_2+p_3)^\beta/m_\rho^2}
	{(p_2+p_3)^2- m_\rho^2 +i \Gamma_\rho m_\rho}
\end{equation}
and 
\begin{equation}
\frac{g^{\alpha\beta}-(p_1+p_3)^\alpha(p_1+p_3)^\beta/m_\rho^2}
	{(p_1+p_3)^2- m_\rho^2 +i \Gamma_\rho m_\rho},
\end{equation}
where $m_\rho$ is the mass of $\rho^0$ and $\Gamma_\rho$ its width.
Alternatively, one can choose propagators in other forms to get better 
approximations.

The $\rho^0\pi^+\pi^-$ vertexes must be vectors if parity is conserved, 
which can be read out from the list of Sec.~\ref{sec:3v}:
\begin{eqnarray}
\Gamma_{\rho\pi^+\pi^-}^\beta (p_2,p_3) & = &
	c(p_2\cdot p_3) \left(p_2-p_3 \right)^\beta,\\
\Gamma_{\rho\pi^+\pi^-}^\beta (p_1,p_3) & = &
	c(p_1\cdot p_3) \left(p_1-p_3 \right)^\beta.
\end{eqnarray}
The $a_1\pi^+\rho^0$ vertexes must be tensors,
\begin{eqnarray}
\Gamma_{a_1\pi^+\rho}^{\mu\alpha}(p_1, p_2+p_3) & = &
  c^{'}_1(p_2\cdot p_3) g^{\mu\alpha}
+ c^{'}_2(p_2\cdot p_3) p_1^\mu p_1^\alpha, \\
\Gamma_{a_1\pi^+\rho}^{\mu\alpha}(p_2, p_1+p_3) & = &
  c^{'}_1(p_1\cdot p_3) g^{\mu\alpha}
+ c^{'}_2(p_1\cdot p_3) p_2^\mu p_2^\alpha.
\end{eqnarray}
Combining all of these together, we find the resonance part of the amplitude 
to be
\begin{equation}
\begin{array}{rcrl}
{\cal M}_\lambda^{(res)} & = &
e_\mu(p,\lambda)\{ &
\;\;\;D_{23}[\frac{3}{2}c_1(s_{23})
	+\frac{1}{2}p_1\cdot (p_2-p_3)c_2(s_{23})]
(p_1+p_2)^\mu\\
& & &
+D_{23}[-\frac{1}{2}c_1(s_{23})
	+\frac{1}{2}p_1\cdot (p_2-p_3)c_2(s_{23})]
(p_1-p_2)^\mu\\
& & &
+D_{13}[\frac{3}{2}c_1(s_{13})
	+\frac{1}{2}p_2\cdot (p_1-p_3)c_2(s_{13})]
(p_1+p_2)^\mu\\
& & &
+D_{13}[-\frac{1}{2}c_1(s_{13})
	+\frac{1}{2}p_2\cdot (p_1-p_3)c_2(s_{13})]
(p_2-p_1)^\mu \}.
\end{array}
\end{equation}
Here $p_3^\mu \simeq - (p_1+p_2)^\mu$ have been used, and
\begin{eqnarray}
s_{23} & = & (p_2 + p_3)^2,\\
s_{13} & = & (p_1 + p_3)^2,
\end{eqnarray}
\begin{eqnarray}
D_{23} & = & \frac{1}{s_{23}-m_\rho^2+i \Gamma_\rho m_\rho},\\
D_{13} & = & \frac{1}{s_{13}-m_\rho^2+i \Gamma_\rho m_\rho}.
\end{eqnarray}
It is easy to see
\begin{eqnarray}
p_1\cdot (p_2-p_3) & = & 
\frac{m_{a_1}^2  + 9 m_\pi^2  - 2 (2 s_{13} + s_{23})}{2},\\
p_2\cdot (p_1-p_3) & = & 
\frac{m_{a_1}^2  + 9 m_\pi^2  - 2 (s_{13} + 2 s_{23})}{2}.
\end{eqnarray}

After redefinition of $c_1$, $c_2$ and $b_2$, the covariant helicity 
amplitude becomes
\begin{equation}
\begin{array}{rcl}
{\cal M}_\lambda & \equiv & {\cal M}_\lambda^{(b)} + {\cal M}_\lambda^{(res)}\\
& = & e_\mu(p,\lambda)(p_1+p_2)^\mu \times \\
& & \{
\;\;b_1\left(s_{13}+s_{23}, (s_{13}-s_{23})^2\right)\\
& & 
\;+ 3c_1(s_{23}) D_{23} 
+ 3c_1(s_{13}) D_{13} \\
& & 
\;+ [m_{a_1}^2  + 9 m_\pi^2  - 2 (2 s_{13} + s_{23})] c_2(s_{23}) D_{23} \\
& & 
\;+ [m_{a_1}^2  + 9 m_\pi^2  - 2 (s_{13} + 2 s_{23})] c_2(s_{13}) D_{13} \} \\
& & +e_\mu(p,\lambda)(p_1-p_2)^\mu \times \\
& & \{
\;\; [s_{13}-s_{23}] 
b_2\left(s_{13}+s_{23}, (s_{13}-s_{23})^2\right) \\
& &
\;-c_1(s_{23})D_{23} + c_1(s_{13})D_{13}\\
& &
\;+[m_{a_1}^2  + 9 m_\pi^2  - 2 (2 s_{13} + s_{23})] c_2(s_{23})D_{23} \\
& &
\;-[m_{a_1}^2  + 9 m_\pi^2  - 2 (s_{13} + 2 s_{23})] c_2(s_{13})D_{13} \}.
\end{array}
\end{equation}

The background amplitude will not give a
flat distribution in the Dalitz plot of the three pions:
\begin{equation}
\begin{array}{rcl}
\sum\limits_{\lambda}|{\cal M}_\lambda^{(b)}|^2 & = &
-|b_1|^2 (p_1+p_2)^2 - |b_2|^2 (p_1-p_2)^2 [(p_1-p_2)\cdot p_3]^2\\
& &
+\frac{1}{m_{a_1}^2} \{
\;\;|b_1|^2[p\cdot (p_1+p_2)]^2 
+ |b_2|^2[p\cdot (p_1-p_2)]^2 [(p_1-p_2)\cdot p_3]^2\\
& & \;\;\;\;\;\;\;\;\;\;
+(b_1^*b_2+b_1b_2^*)
  [p\cdot (p_1+p_2)][p\cdot (p_1-p_2)][(p_1-p_2)\cdot p_3] \}.
\end{array}
\end{equation}
In fact for any process involving particles with non-zero spins,
the background distributions are not flat.
If we do not include such background(1PI) terms, those resonances
terms we have considered might just simulating the background distributions.
Let's take the process $a_1\rightarrow \pi^+\pi^+\pi^-$ as an example.
We can not say that we have seen $\rho\prime$ or some other resonances, if
the background term ${\cal M}_\lambda^{(b)}$ is not considered when we 
fit data. Particularly, this {\em is} the case for resonances far off shell
or with large widths. {\bf One must include background terms}. A resonance
in a process can be taken as well established under two conditions:
(1) We must include the background terms in amplitudes with and without such a
resonance. (2) The amplitudes including the resonance  significantly 
improve the best fit to  experimental data, comparing to those ignoring it.

\section{Summary}
The main results of this paper are summarized bellow:

A list of general 3-leg effective vertexes for bosons is given,
with kinematic singularities carefully avoided. 
Space reflection symmetry demand effect vertexes to be tensors or
pseudo-tensors depending on spin-parities of external lines. 
Mixing of tensor and pseudo-tensor vertexes always means violation
of parity conservation. Boson symmetry require that effective vertexes 
take the special form given in Sec.\ref{sec:bs}. 
The requirement of parity conservation and boson symmetry leads to 
selection rules. 
These results are needed when we construct phenomenological models
or write amplitudes to fit data of high energy experiments.

Helicity amplitudes in laboratory frame are related to those in
center of mass frame by Wigner rotations. 
For two-body decays, it is possible to write the 
explicit expressions of covariant helicity amplitudes in a concise
form.

S-matrixes for processes involving more than three particles can be divide
into 1PI parts and one-particle reducible parts, or in another words, 
backgrounds and resonances. We emphasize that such background
terms are important when one try to extract meaningful information 
on resonances. This is especially the case if the width of the resonances
are large, or the resonances are far off shell. 

Constraints of gauge invariance on effective vertexes will be the content 
of another paper.

\section*{Acknowledgement}
Jie-Jie Zhu wishes to thank Prof. Yu-Can Zhu and Dr. Liao-Yuan Dong 
for discussions.
This work is partially supported by National Founds of China 
through C. N. Yang, Funds of IHEP of China and the Grant 
LWTZ-1298 of Chinese Academy of Science.

\end{document}